\documentstyle[amsmath,aps,prb,multicol,epsfig,floats,axodraw,amssymb,eqsecnum,graphics]{revtex}
\begin{document}

\twocolumn[\hsize\textwidth\columnwidth\hsize\csname @twocolumnfalse\endcsname
\title{The two-dimensional random-bond Ising model, free fermions and the network model}
\author{F. Merz and J. T. Chalker}
\address{Theoretical Physics, University of Oxford, 1 Keble Road,
Oxford  OX1 3NP, UK}
\date{\today}
\maketitle

\begin{abstract}
We develop a recently-proposed mapping of the two-dimensional Ising model with 
random exchange (RBIM), via the transfer matrix, to a network model for a disordered system of non-interacting fermions.
The RBIM transforms in this way to a localisation problem belonging to one of a set of non-standard
symmetry classes, known as class D; the transition between paramagnet and ferromagnet is equivalent to 
a delocalisation transition between an insulator and a quantum Hall conductor.
We establish the mapping as an exact and efficient
tool for numerical analysis: using it, the computational effort required to study
a system of width $M$ is proportional to $M^{3}$, and not exponential in $M$ as with conventional algorithms.
We show how the approach may be used to calculate for the RBIM: the  free energy; 
typical correlation lengths in quasi-one dimension for both the spin and the disorder operators; even powers of spin-spin
correlation functions and their disorder-averages.
We examine in detail the square-lattice, nearest-neighbour $\pm J$ RBIM, in
which bonds are independently antiferromagnetic with probability $p$, and ferromagnetic with probability $1-p$.
Studying temperatures $T\geq 0.4J$, we obtain precise coordinates in the $p-T$ plane for points on the phase boundary between ferromagnet and paramagnet, 
and for the multicritical (Nishimori) point. We demonstrate scaling flow towards the pure Ising fixed point at small $p$, and determine critical
exponents at the multicritical point.
%\end{abstract}
%\vskip 0.2 truein
\end{abstract}
]
%\newpage
\vskip2pc

\section{Introduction}
The two-dimensional Ising model \cite{baxter,mccoywu} has been a basic prototype in the theory
of phase transitions for over half a century. A central factor in its importance
has been its equivalence to a system of non-interacting fermions, as set
out by Schultz, Mattis and Lieb \cite{SML} in their well-known reformulation
of Onsager's solution.
The two-dimensional Ising model has naturally also been a test-bed
for studies of the effect of quenched disorder on phase transitions, and
the equivalence between the spin system and free fermions continues
to hold in the presence of randomness in exchange interactions. In this
paper we build on recent work by Cho and Fisher,\cite{chofisher,chothesis} and by Gruzberg, Read and Ludwig\cite{readlud,GRL} to establish the correspondence in a form suitable for numerical
analysis, and use it to study the square-lattice, random-bond Ising model (RBIM).

The consequences for the two-dimensional Ising model of weak randomness in exchange interactions are rather well understood, following analytical calculations based on the fermionic formulation by Dotsenko and Dotsenko \cite{dotdot} and others:\cite{shalaev,shankar,ludwig} weak disorder is marginally irrelevant in the renormalisation group sense, and the  thermally-driven transition from the paramagnet to the ferromagnet survives with only logarithmic modifications to the critical behaviour of the pure system. By contrast, strong disorder has more dramatic effects. A convenient choice is to consider exchange interactions with fixed magnitude which are independently ferromagnetic or antiferromagnetic, with probabilities $1-p$ and $p$ respectively. In this case, it is known from a variety of approaches\cite{youngsouthern,rammal,nishiorg,nishi,LeDouss,kitavert,singh,OzNishi1,houdayer,morgbin,mcmillan,OzNishi,UenoOz,KitaOgu,queiroz,honecker,Ozeki,KawaRieger} that the Curie temperature is depressed with increasing $p$, reaching zero at a critical disorder strength, $p_c$. Moreover, while the scaling flow at the transition is controlled for small $p$ by the critical fixed point of the pure system, at larger $p$ it is determined by a disorder-dominated multicritical point, known as the Nishimori point.\cite{nishiorg,nishi,LeDouss}

Most numerical studies of the RBIM have used either Monte Carlo simulations \cite{OzNishi1,houdayer} or transfer matrix calculations in a spin basis. \cite{morgbin,mcmillan,OzNishi,UenoOz,KitaOgu,queiroz,honecker} Fermionic formulations of the Ising model nevertheless have two great potential advantages: they can avoid the statistical sampling errors of Monte Carlo simulations; and also, if implemented using the transfer matrix,  they can avoid the exponential growth in transfer matrix dimension with system width that occurs if this matrix is written in a spin basis. Pioneering steps in the first of these directions have been taken by Blackman \cite{blackman} and collaborators,\cite{blackmanplus}
and others,\cite{saulkardar,inoue,Sorensen} using the solution of the two-dimensional Ising model via a Pfaffian\cite{mccoywu} to express statistical-mechanical quantities in terms of spectral properties of the associated matrix. Their work makes a link between the RBIM and localisation problems, since the matrix allied to the Pfaffian is essentially a tight-binding Hamiltonian on the lattice of the underlying Ising model, with random hopping arising from random exchange. An alternative route from the RBIM to a localisation problem has been proposed by Cho and Fisher:\cite{chofisher,chothesis} starting from two copies of the transfer matrix for an Ising model, each expressed in terms of Majorana fermions and combined to form Dirac fermions, they arrive at a version of the network model similar to that introduced as a description for the integer quantum Hall plateau transition,\cite{chalcodd} though with a distinct symmetry. 

Viewed as a localisation problem, the paramagnetic and ferromagnetic phases of the RBIM translate to two insulating phases with Hall conductance differing by one quantum unit, while the Curie transition maps to a version of the quantum Hall plateau transition. This transition, and indeed the insulating phases, belong to a non-standard symmetry class for localisation, classified in work by Altland and Zirnbauer  \cite{AltZirn} and known as class D. The match between behaviour expected in the RBIM and that anticipated for two-dimensional localisation problems in class D has been the subject of recent discussion. \cite{readlud,GRL,bundschu,Senthil,readgreen,bocquet,chalker} A particular difficulty has been to reconcile the fact that, generically, a third, metallic phase is possible in the localisation problem, in addition to the two insulating phases, while the RBIM in two dimensions is expected to display only two phases. The resolution which has emerged \cite{readlud,chalker} is that symmetry alone is not sufficient to determine the phases that appear, and that in the specific disordered conductor equivalent to the RBIM no metallic phase arises.

The work we describe here builds on Cho and Fisher's ideas, which must be extended in several ways to provide a precise and practical treatment of the RBIM. First, the approach described in Ref.\,\onlinecite{chofisher} proceeds from the RBIM via a continuum limit, which is rediscretised to obtain a network model. In order to find an explicit relationship between parameters in the two systems, it is necessary instead to carry out the mapping directly on a lattice model. Doing so, as described by Cho in her thesis\cite{chothesis} and by Gruzberg, Read and Ludwig in Refs.\,\onlinecite{readlud} and \onlinecite{GRL}, one arrives at a network model different in detail to that studied in Ref.\,\onlinecite{chofisher}, and with different behaviour.\cite{chalker}
Second, a proper treatment of the RBIM in cylindrical geometry requires an appropriate choice of boundary conditions in the network model, which has not previously been considered. Third, to calculate thermodynamic quantities, typical correlation lengths, spin and disorder correlation functions for the RBIM using the network model formulation, it is necessary to map from fermions back to spins, as outlined in Refs.\,\onlinecite{readlud} and \onlinecite{GRL} and as we describe here. A feature of interest which emerges from our analysis is a topological distinction between the paramagnetic and ferromagnetic phases as represented in terms of fermions, similar to that discussed recently for other systems from symmetry class D.\cite{readlud,motrunich} Finally, an important technical aspect of the work we present here is that numerical transfer matrix calculations for localisation problems in the symmetry class we are concerned with require for numerical stability a modification of the standard algorithm, as first discussed in Ref\,\onlinecite{chalker}.

As a numerical approach to the RBIM, the method we describe has two main limitations. One arises because the Dirac fermions of the network model are built from two copies of an Ising model. As a result, it turns out to be straightforward to calculate even powers of spin correlations functions, and their disorder-averages, but not practical to calculate odd powers. The other stems from the fact that Boltzmann factors which enter the network model become large at low temperatures, making the zero temperature limit inaccessible. 

The remainder of the paper is organised as follows: In Sec.\,\ref{tm} and Sec.\,\ref{ntwk} we outline the Jordan-Wigner fermionisation of the spin transfer 
matrix and the mapping to a network model. In Sec.\,\ref{LE} we discuss boundary conditions across the system in network model language 
and the subsequent rules for constructing the spin transfer matrix from the fermion transfer matrix. In Sec.\,\ref{NSC} and Appendix\,\ref{round} we review the numerical algorithm that we employ in the network model transfer matrix calculations and set out how statistical-mechanical quantities are obtained from the fermion description. In Sec.\,\ref{rbim} we present numerical results on the
$\pm J$ RBIM. The system sizes we study (transverse width $M=8-256$ spins) 
are significantly larger than what was previously possible.
We focus on critical behaviour at the Nishimori point, for which we determine the coordinate $p_{c}=0.1093\pm 0.0002$.
We calculate the critical exponents $\nu$ and $\nu_{T}$, describing the divergence of the correlation length as the Nishimori point is approached 
along the Nishimori line and the phase boundary, respectively. Using large system sizes we find $\nu=1.50\pm 0.03$, in disagreement with previous estimates,\cite{singh,honecker} and, with wider confidence limits, $\nu_{T}=4.0\pm 0.5$.

\section{Transfer Matrix}\label{sec.TM}
\subsection{Ising model transfer matrix}\label{tm}
\begin{figure}[ht]
\begin{center}
\epsfig{file=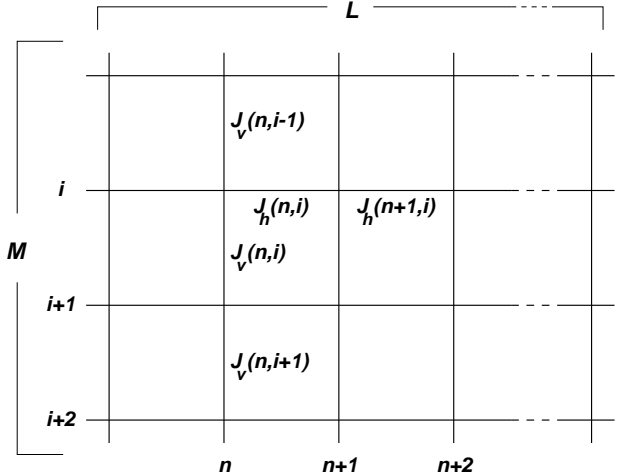,width=3.2in}\vspace{0.2cm}\\
\caption{\small For the square lattice Ising model we adopt the convention that a pair
$(n,i)$ labels one spin with two associated bonds, one horizontal (to the right) and one vertical (downwards).\label{ising}}
\end{center}
\end{figure}

We consider the nearest neighbour Ising model on a square lattice in two dimensions. 
The partition function $Z$ for a such a system on a strip of length $L$ and width $M$ can be written \cite{baxter} in terms of a product of 
transfer matrices. Introducing integer coordinates, $n$ and $i$, as illustrated in Fig.\ref{ising}, one has 
\begin{equation}\label{partition}
Z= {\cal A}\ \mbox{Tr}\left[\hat{T}_{1}\hat{T}_{2}\cdots\hat{T}_{n}\cdots \hat{T}_{L}\right],
\end{equation}
with $\hat{T}_{n}=\hat{V}_{n}\times \hat{H}_{n}$ and
\begin{equation}\label{transfer}
\begin{array}{rl}
\hat{H}_{n}= & \exp\left(-\sum_{i=1}^{M}[\kappa^{*}_{n,i}\sigma^{z}_{i}]\ \right) \vspace{0.3cm}\\
\hat{V}_{n}= & \exp\left(\sum_{i=1}^{M}[ \kappa_{n,i}\sigma^{x}_{i}\sigma^{x}_{i+1}]\ \right), \vspace{0.3cm}\\
\end{array}
\end{equation}
where the $\sigma$'s are Pauli matrices and
\begin{equation}\label{KW}
\begin{array}{c}
\kappa_{n,i}=\beta J_{v}(n,i), \vspace{0.4cm}\\
\kappa^{*}_{n,i}=-\frac{1}{2}\mbox{ln}\ [\mbox{tanh}\ \beta J_{h}(n,i) ],\vspace{0.4cm}\\
{\cal A} =\prod_{n=1}^{L}\prod_{i=1}^{M}\ \sqrt{2} \left[ \mbox{sinh}\ 2\kappa^{*}_{n,i}\right]^{-\frac{1}{2}}.
\end{array}
\end{equation}
Here, $\kappa_{n,i}$ is the reduced coupling strength at inverse temperature $\beta$ between
the $i$th and $(i+1)$th spin in the vertical
direction of Fig.\ref{ising} on the $n$th slice,  and $\kappa^{*}_{n,i}$ is the Kramers-Wannier dual value of the corresponding bond strength in the
horizontal direction. For the rest of the paper the labels $v$ and $h$ on the bond strengths
are redundant, since all horizontal bond strengths (and only those) appear as dual values, identified with an asterix. 
We take $\sigma^{x}_{M+1}\equiv\sigma^{x}_{1}$ in Eq.\,(\ref{transfer}) so that boundary conditions across the strip are controlled by the 
set of interactions strengths $\kappa_{n,M}$. For convenience we introduce the notation  $\hat{T}(k,l)\equiv\prod_{n=k}^{l}\hat{T}_{n}$, and 
for brevity we use $\hat{T}$ to denote either $\hat{T}(k,l)$ or $\hat{T}_{n}$.

Following Schultz, Mattis and Lieb \cite{SML} the operators $\hat{H}_{n}$ and $\hat{V}_{n}$ can be written, using
the Jordan-Wigner-transformation, as functions of fermionic operators. Introducing the fermion annihilation and creation operators 
$C_{i}$ and $C_{i}^{\dagger}$, the spin operators become
%\begin{equation}\label{JW}
%\begin{array}{l}
%\sigma^{+}_{i}\equiv\sigma^{x}+i\sigma^{y}=
%\exp (i\pi\sum_{j=1}^{i-1}C^{\dagger}_{j}C_{j})\ %C^{\dagger}_{i},\vspace{0.2cm}\\
%\sigma^{-}_{i}\equiv\sigma^{x}-i\sigma^{y}=
%\exp (i\pi\sum_{j=1}^{i-1}C^{\dagger}_{j}C_{j})\ C_{i},\vspace{0.2cm}\\
%\sigma^{z}_{i}=2C_{i}^{\dagger}C_{i}-1
%=-\exp (i\pi C^{\dagger}_{i}C_{i}).\vspace{0.2cm}\\
%\end{array}
%\end{equation}
\begin{equation}\label{JW}
\begin{array}{l}
\sigma_i^{x}=
\exp(i\pi\sum_{j=1}^{i-1}C^{\dagger}_{j}C_{j})\ (C^{\dagger}_{i}+C_i)\,,\vspace{0.2cm}\\
%\end{equation}
%and
%\begin{equation}
\sigma^{z}_{i}=2C_{i}^{\dagger}C_{i}-1\,.
%=-\exp (i\pi C^{\dagger}_{i}C_{i})\,.
\end{array}
\end{equation}
After Jordan-Wigner transformation, $\hat{H_{n}}$ and $\hat{V}_{n}$ read
\begin{equation}\label{Ctrans}
\begin{array}{ll}
\hat{H}_{n} & = \exp\left(-2\sum_{i=1}^{M}\ \kappa^{*}_{n,i}[C^{\dagger}_{i}C_{i}-\frac{1}{2}]\right)\,,
\vspace{0.4cm}\\
\hat{V}_{n} & = \exp\Big(\sum_{i=1}^{M-1}\ \kappa_{n,i}[C^{\dagger}_{i}-C_{i}][C^{\dagger}_{i+1}+
C_{i+1}] \vspace{0.3cm}\\
& \quad  -\kappa_{n,M}e^{i\pi N_{C}}[C^{\dagger}_{M}-C_{M}]
[C^{\dagger}_{1}+C_{1}]\Big),
\end{array}
\end{equation}
with $N_{C}=\sum_{i=1}^{M}\ C^{\dagger}_{i}C_{i}$, the number operator.
A familiar feature of the transfer
matrix in fermionic language is that it does
not conserve  $N_{C}$, since $\hat{V}_{n}$ includes terms which create and annihilate fermions in pairs. 
Such a structure is reminiscent of Bogoliubov-de Gennes Hamiltonians 
arising in the mean-field description of superconductors.
It has the consequence that,
to diagonalise the transfer matrix for a translationally invariant Ising model, one uses Fourier transformation followed by Bogoliubov transformation \cite{}.
For the RBIM without translational invariance, the transformation
that diagonalises the transfer matrix is disorder-dependent, and one must follow a different route to make progress. 

In place of diagonalisation, the objective for the RBIM is to write the
transfer matrix in terms of Dirac fermions whose number is conserved under its action. The necessary steps are well-established \cite{zubitz,shankar} and have been set out in the present context by Cho and Fisher,\cite{chofisher} Cho,\cite{chothesis} and
Gruzberg, Read and Ludwig \cite{GRL}. 
First, because of the form of Eq.\,(\ref{Ctrans}), it is natural to decompose the complex (Dirac) fermions into real and imaginary parts, introducing real
(Majorana) fermions $\xi_{C}$ and $\eta_{C}$.
Suppressing the site index one can write
\begin{equation}
C=\frac{1}{\sqrt{2}}[\xi_{C}-i\eta_{C}]\,,\qquad C^{\dagger}=\frac{1}{\sqrt{2}}[\xi_{C}+i\eta_{C}]\,,
\end{equation}
where $\xi_{C}$ and $\eta_{C}$ anticommute and satisfy $\xi_{C}^{\dagger}=\xi_{C}$, $\eta_{C}^{\dagger}=\eta_{C}$ and $\{\xi_{Ci},\xi_{Cj}\}=\{\eta_{Ci},\eta_{Cj}\}=\delta_{ij}$.
Next, in order to return to Dirac fermions, one introduces
a second, identical copy of the Ising 
model. 
We represent the second copy using the Dirac fermions 
$D$ and $D^{\dagger}$, in analogy to the $C$ and $C^{\dagger}$, and employ the Majorana decomposition $D=[\xi_{D}-i\eta_{D}]/\sqrt{2}$ and
$D^{\dagger}=[\xi_{D}+i\eta_{D}]/\sqrt{2}$. This provides new ways to recombine the Majorana fermions. Of the various alternatives, 
consider in particular the Dirac fermions $f=[\xi_{C}+i\xi_{D}]/\sqrt{2}$ and $g=[\eta_{D}-i\eta_{C}]/\sqrt{2}$, which we choose to yield real 
coefficients later on. Again suppressing the site index, this transformation may be summarised by 
\begin{equation}\label{summtr}
\begin{array}{c}
C=\frac{1}{2}[f+f^{\dagger}+g-g^{\dagger}]\,,\vspace{0.3cm}\\
D=\frac{i}{2}[f^{\dagger}-f-g-g^{\dagger}]\,,\vspace{0.3cm}\\
\end{array}
\end{equation}
and its inverse
\begin{equation}
\begin{array}{c}
f=\frac{1}{2}[C+C^{\dagger}+iD+iD^{\dagger}]\,,\vspace{0.3cm}\\
g=\frac{1}{2}[C-C^{\dagger}+iD-iD^{\dagger}]\,.
\end{array}
\end{equation}
As an aside, we note that the Jordan-Wigner transformation applied to two copies of the Ising model does not by itself generate the correct commutation relations
between pairs of spin operators $\sigma^{x}$ taken one from each copy. To ensure these commutation relations one should in addition introduce Klein factors. Since the Klein factors ultimately have no effect on the equations we present, we omit them throughout this paper.

For the doubled system, we are concerned with the transfer matrix products
(suppressing the slice index) $\hat{H}_C\hat{H}_{D}$ and
$\hat{V}_{C}\hat{V}_{D}$. The value of the transformation Eq.\,(\ref{summtr}) 
is that it reduces these products to the simple forms
\begin{equation}\label{transfer2}
\begin{array}{lcl}
\hat{H}_{C}\hat{H}_{D} & =  & \exp\left(-2\sum_{i=1}^{M}\ \kappa^{*}_{n,i}[g^{\dagger}_{i}f_{i}+f^{\dagger}_{i}g_{i}]\right)\vspace{0.3cm}\\
\hat{V}_{C}\hat{V}_{D} & = & \exp\left(2\sum_{i=1}^{M-1}\ \kappa_{n,i}[g^{\dagger}_{i}f_{i+1}+f^{\dagger}_{i+1}g_{i}]+ 
 B\ \right)\,, 
\end{array}
\end{equation}
where the boundary term $B$ is
\begin{equation}\label{BT}
\begin{array}{l}
B=-\kappa_{n,M}\left[ (e^{i\pi N _{C}}+e^{i\pi N_{D}})[g^{\dagger}_{M}f_{1}+f^{\dagger}_{1}g_{M}]\right. \vspace{0.2cm}\\
\qquad \left. +(e^{i\pi N_{C}}-e^{i\pi N_{D}})[g^{\dagger}_{M}f^{\dagger}_{1}+f_{1}g_{M}]\right].
\end{array}
\end{equation}
This process of doubling the degrees of freedom and rewriting them locally as fermions, in order to remove terms which are not particle conserving,
may be viewed as a local Bogoliubov transformation.

The boundary term $B$ contains the two boundary operators 
\begin{equation}\label{BO}
{\cal B}^{\pm}=e^{i\pi N _{C}}\pm e^{i\pi N_{D}}.
\end{equation} 
These operators commute with the transfer matrix as a consequence of ${\rm Z}_{2}$ symmetry:
for a single system, say $C$, one can identify two invariant subspaces, distinguished by the behaviour 
of vectors within the subspace under the operation $\hat{R}_{C}$ which reverses the orientation of a complete row of spins.\cite{baxter} Specifically,
\begin{equation}
\begin{array}{c}
\hat{R}_{C}=\prod_{i}\sigma^{z}_{iC}\,\,,\vspace{0.3cm}\\
\hat{R}_{C}\sigma^{x}_{jC}\hat{R}_{C}=-\sigma^{x}_{jC}\,\, 
\end{array}
\end{equation}
for all $j$, and $\hat{R}_{C}^{2}=1$. Introducing the corresponding operator $\hat{R}_{D}$ for the $D$ system and assuming the total number of spins across the
strip to be even, one finds that the boundary operators
are simply ${\cal B}^{\pm}=\hat{R}_{C}\pm\hat{R}_{D}$. Since both $\hat{R}_{C}$ and $\hat{R}_{D}$ commute with
the transfer matrix, four invariant subspaces arise naturally from
$[\hat{R}_{C}=\pm 1]\otimes [\hat{R}_{D}=\pm 1]$. Using obvious notation, $\hat{T}_{C}\hat{T}_{D}$ may then be presented schematically in the block-diagonal form 
\begin{equation}\label{sectors}
\hat{T}_{C}\hat{T}_{D}=
\left(\begin{array}{cccc}
++ & 0 & 0 & 0 \vspace{0.2cm}\\
0 & -+ & 0 & 0 \vspace{0.2cm}\\
0 & 0 & +- & 0 \vspace{0.2cm}\\
0 & 0 & 0 & --
\end{array}\right).
\end{equation}
Thus the Fock space associated with the $C$ and $D$ fermions
can be divided into four subspaces according to the parity of $N_{C}$ and $N_{D}$. In two of them, for which ${\cal B}^{-}=0$, the number
of $f$ and $g$ fermions is conserved under the action of $\hat{T}$.

\subsection{Network model interpretation}\label{ntwk}

The conservation of the Dirac fermions $f$ and $g$ under the action of the transfer matrix operator makes it possible to go from a second-quantised description to a first-quantised form. Moreover, just as the second-quantised form has SO(2) symmetry,\cite{GRL} one finds that the first-quantised form may be interpreted as the transfer matrix for a scattering problem, because it fulfills the requirements arising from unitarity of the scattering matrix. Specifically, the first-quantised form represents a 
network model, in which non-interacting $f$ and $g$ fermions propagate on directed links of a lattice. The fermions scatter at nodes, where two incoming links and two outgoing links meet. In this way, the nodes of the network model take the place of bonds in the Ising model. A correspondence of this type was set out first by Cho and Fisher\cite{chofisher} and subsequently refined by Cho,\cite{chothesis}  who pointed out that the network model studied numerically in Ref.\onlinecite{chofisher} is equivalent to an Ising model in which some exchange couplings are imaginary, while the RBIM itself is represented by a different network model. In this subsection we review these ideas.

The identification of the first-quantised form of $\hat{T}$ makes use of a general equivalence between first- and second-quantised versions of linear transformations. Consider in a Hilbert space of dimension $N$ a linear transformation of single-particle wavefunctions, represented in a certain basis by an $N\times N$ matrix with elements $(\exp G)_{ij}$. Introducing in the same basis fermion creation and annihilation operators, $\alpha^{\dagger}_{i}$ and $\alpha_{i}$, the second-quantised representation of this transformation is $\exp [\alpha^{\dagger}_{i}G_{ij}\alpha_{j}]$.
To apply this equivalence to the transfer matrix $\hat{T}$, 
let the $N\equiv2M$ fermion annihilation operators be $[\alpha_{1},\cdots,\alpha_{2M}]=[f_{1},\cdots,f_{M},g_{1},\cdots,g_{M}]$.
In the ${\cal B}^{-}=0$ subspaces, the transfer matrix of the RBIM has the canonical form
\begin{equation}\label{gentrans}
\hat{T}_{C}\hat{T}_{D}=\exp [\alpha^{\dagger}_{i}G_{ij}\alpha_{j}],
\end{equation}
and can be represented equivalently by the $2M\times 2M$ matrix $T$, with elements $T_{ij}=(\exp G)_{ij}$, as a transformation of single-particle states. Thus the action of the operator $\hat{T}$ on a Slater determinant is replicated by the action of the matrix $T$ on the orbitals entering the determinant.
In the following we use notation for the matrix $T$ corresponding to that introduced for the operator $\hat{T}$: $T_{n}$ denotes the transfer matrix for the n-th slice of the system, $T(k,l)$ indicates a product and $T$ is shorthand for either.

\begin{figure}[ht]
\begin{center}
\epsfig{file=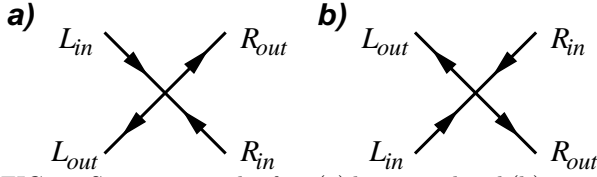,width=3.1in}
\caption{\small Scattering nodes for: (a) horizontal and (b) vertical bonds.\label{node}}
\end{center}
\end{figure}

While knowledge of the single-particle form of $T$ is enough by itself for efficient numerical calculations, physical interpretation within this framework of the RBIM as a localisation problem depends on the fact that $T$ is a pseudo-orthogonal matrix. In consequence, it can be viewed as the transfer matrix for a scattering problem in which flux is conserved. 
In order to see that this is indeed the case, consider the basic building blocks of the transfer matrix for one column of sites in the doubled Ising model.
The two factors, $\hat{H}_{D}\hat{H}_{C}$ and $\hat{V}_{D}\hat{V}_{C}$, appearing in Eq.\,(\ref{transfer}) each consist of products 
of $M$ commuting operators. Every such operator represents a single bond of the Ising model and involves only one pair of $f$ and $g$ fermions. Schematically, a horizontal bond gives rise to $\exp(-2\kappa^{*}[g^{\dagger}f+f^{\dagger}g])$, which is replaced in a first-quantised treatment by the $2\times 2$ matrix $h\equiv \exp (-2\kappa^{*} \sigma^{x})$, while a vertical bond yields $\exp(2\kappa[g^{\dagger}f+f^{\dagger}g])$, which is replaced by $v\equiv \exp (2\kappa \sigma^{x})$. To arrive at a scattering problem, the $f$ fermions are regarded (arbitrarily) as right-movers, and the $g$ fermions as left-movers. 
Then the matrices $h$ and $v$ are transfer matrices for nodes of the network model. They relate flux amplitudes, $L_{in}$ and $L_{out}$, to the amplitudes 
$R_{in}$ and $R_{out}$, appearing either side of a node as illustrated in Fig.\,\ref{node}.
In algebraic terms, we have for horizontal bonds the equation
\begin{equation}\label{nodes-h}
\left( \begin{array}{c} R_{out} \vspace{0.2cm}\\ R_{in}\end{array}\right) =
\left(
\begin{array}{cc}
\cosh2\kappa^{*} & -\sinh2\kappa^{*} \vspace{0.2cm}\\
-\sinh2\kappa^{*} & \cosh2\kappa^{*}
\end{array}\right)
\left( \begin{array}{c} L_{in} \vspace{0.2cm}\\ L_{out}\end{array}\right)
\end{equation}
and for vertical bonds the equation
\begin{equation}\label{nodes-v}
\left( \begin{array}{c} R_{in} \vspace{0.2cm}\\ R_{out}\end{array}\right) = 
\left(
\begin{array}{cc}
\cosh2\kappa & \sinh2\kappa \vspace{0.2cm}\\
\sinh2\kappa & \cosh2\kappa
\end{array}\right)
\left( \begin{array}{c} L_{out} \vspace{0.2cm}\\ L_{in}\end{array}\right)\,.
\end{equation}
Flux conservation follows from the relations $\sigma^{z}h^{\dagger}\sigma^{z}=h^{-1}$ and $\sigma^{z}v^{\dagger}\sigma^{z}=v^{-1}$. 

\begin{figure}[ht]
\begin{center}
\epsfig{file=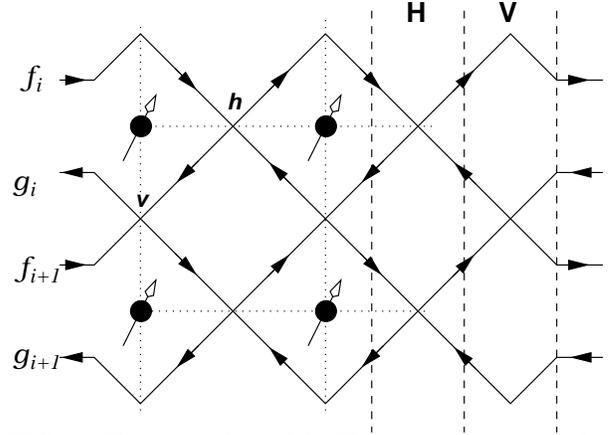,width=3.1in}
\caption{\small The network model. Flux propagates on links in the direction
indicated by arrows. The transfer matrix relates flux amplitudes carried by links on the right to those on the left. Nodes arising from
single rows of vertical bonds and horizontal bonds in the Ising model are indicated by $V$ and $H$, respectively. Two particular nodes are labelled
by $h$ and $v$. Four sites of the Ising model are also shown with exchange interactions as dotted lines.\label{network}}
\end{center}
\end{figure}
The network model as a whole is illustrated in Fig.\,\ref{network}.
It has the same structure as the U(1) network model, introduced to describe localisation in the context of the integer quantum Hall effect.\cite{chalcodd}
Directed links form plaquettes, each with a definite sense of circulation, which is alternately clockwise and anticlockwise on successive squares. Disorder appears in the U(1) network model in the form of quenched random phases associated with links. By contrast, for the RBIM randomness enters only through the scattering parameters, $2\kappa$ and $2\kappa^{*}$, associated with nodes. An antiferromagnetic vertical bond leads to a negative node parameter, $\kappa$. An antiferromagnetic horizontal bond, however, gives rise to a complex $\kappa^{*}$, since from Eq.\,(\ref{KW})
\begin{equation}
(-|\kappa|)^{*}=|\kappa|^{*}+i\pi/2\,,
\end{equation}
generating an overall minus sign for $h$. The sign is accompanied by a minus sign as a factor in the coefficient ${\cal A}^{2}$, defined in Eq.\,(\ref{KW}).

The form of this disorder determines the symmetry class to which this network model belongs in the classification introduced by Altland and Zirnbauer\cite{AltZirn}. Specifically, Hamiltonians $H$ belonging to class D have, in a suitable basis, the property that $H^*=-H$, so that $H$ is pure imaginary. Adapting this defining relation to a network model, one supposes that propagation on the network  is generated by a time-evolution operator for unit time-step, $\exp(iH)$. For class D, this evolution operator is real, so that scattering phase factors may take only the values $\pm 1$, as is indeed the case for the RBIM. In detail, a single antiferromagnetic bond (either horizontal or vertical) introduces phases of  $\pi$ for propagation around both the anticlockwise plaquettes that meet at the corresponding node, compared to the phases in the purely ferromagnetic model. Other choices of randomness belonging to the same symmetry class are of course possible. Cho and Fisher\cite{chofisher} investigated a model in which the transfer matrices at all nodes are of the type given in Eq.\,(\ref{nodes-v}), with randomness in the sign of $\kappa$, while other authors\cite{readlud,chalker} have studied a model in which scattering phase factors of $\pm 1$ are associated independently with links rather than nodes. Strikingly, each of these different choices leads to very different localisation properties in the network model.\cite{readlud,chalker} 

Combining the $2\times 2$ transfer matrices, $h$ or $v$, for each node, one arrives at the $2M\times 2M$ transfer matrix $T$ for the system as a whole. Flux conservation guarantees that $T$ may be factorised as
\begin{equation}\label{T1}
T=\left(\begin{array}{cc}
W_{L} & 0 \vspace{0.3cm}\\
0 & V^{T}_{L} \end{array} \right) 
\left(\begin{array}{cc}
\cosh(\epsilon L) & \sinh(\epsilon L) \vspace{0.3cm}\\
\sinh(\epsilon L) & \cosh(\epsilon L) \end{array} \right) 
\left(\begin{array}{cc}
W^{T}_{R} & 0 \vspace{0.3cm}\\
0 & V_{R} \end{array} \right),
\end{equation}
where components in the basis are ordered so that the amplitudes for propagation in one direction constitute the first $M$ entries
of the vectors on which $T$ acts, and those for propagation in the opposite direction make up the remaining $M$ entries.
Here, the $M\times M$ matrices, $W_L$, $W_R$, $V_L$ and $V_R$ are for a general localisation problem unitary matrices, and for the Ising model orthogonal matrices, since in that case every element of the transfer matrix is real. The $M\times M$ matrix $\epsilon$ is real, positive and diagonal.
It is convenient to rewrite Eq.\,(\ref{T1}) in the form
\begin{equation}\label{PD}
T=\frac{1}{2}\left(\begin{array}{cc}
W_{L} & -W_{L} \vspace{0.3cm}\\
V^{T}_{L} & V^{T}_{L} \end{array} \right) 
\left(\begin{array}{cc}
e^{\epsilon L} & 0 \vspace{0.3cm}\\
0 & e^{-\epsilon L} \end{array} \right) 
\left(\begin{array}{cc}
W^{T}_{R} & V_{R} \vspace{0.3cm}\\
-W^{T}_{R} & V_{R} \end{array} \right),
\end{equation}
where the diagonal elements of $\exp(\pm \epsilon L)$ are the singular values of $T$.
For a random system of length $L$, the exponents $\epsilon L$ are ${\cal O}(L)$, with sample-to-sample fluctuations which are ${\cal O}(L^{1/2})$. From Oseledec's theorem, the average $\epsilon$ tends
to a limit, $\mbox{diag}(\epsilon_{1},\epsilon_{2},\cdots,\epsilon_{M})$, for large $L$, where $\epsilon_1 \leq \epsilon_2 \leq \ldots \epsilon_M$ are the Lyapunov exponents characterising the network model.
 
It is useful also to express Eq.\,(\ref{PD}) in second-quantised notation. Writing the left and right orthogonal matrices in 
terms of the Hermitian $2M \times 2M$ matrices $A_{L}$ and $A_{R}$, defined by
\begin{equation}
\begin{array}{c}
\exp [-i A_{L}]\equiv\frac{1}{\sqrt{2}}\left(\begin{array}{cc}
W_{L} & -W_{L} \vspace{0.3cm}\\
V^{T}_{L} & V^{T}_{L} \end{array} \right),\vspace{0.3cm}\\ 
\exp [i A_{R}]\equiv\frac{1}{\sqrt{2}}\left(\begin{array}{cc}
W^{T}_{R} & V_{R} \vspace{0.3cm}\\
-W^{T}_{R} & V_{R} \end{array} \right),
\end{array}
\end{equation}
the transfer matrix for the doubled Ising model takes the form (within the subspaces with ${\cal B}^- = 0$)
\begin{equation}\label{2ndQT}
\begin{array}{rl}
\hat{T}_{C}\hat{T}_{D}= & \exp [-i\alpha^{\dagger}_{i}A_{Lij}\alpha_{j}]\times \exp [\alpha^{\dagger}_{i}\alpha_{i}[\sigma^{z}\otimes \epsilon L]_{ii}]\vspace{0.3cm}\\
& \qquad \times \exp [i\alpha_{i}^{\dagger}A_{Rij}\alpha_{j}]\,.
\end{array}
\end{equation}

\subsection{Lyapunov exponent spectrum}\label{LE}

In this subsection we discuss some aspects of the mapping between the RBIM and the network model that have not been considered in previous work. These stem from the fact that, under the Jordan-Wigner transformation, different boundary conditions arise in $\hat{T}$ according to the parity of the fermion numbers $N_C$ and $N_D$ (see Eq.\,(\ref{BT})). Full information on sectors of both parities is contained in the results of network model calculations for the subspaces denoted $++$ and $--$ in Eq.\,(\ref{sectors}). To make use of this information it is necessary establish how the Lyapunov exponents of the spin transfer matrix are related to those of the network model. A crucial step is to be able to identify the parity of left and right vectors of $\hat{T}$ when these are written in terms of the $f$ and $g$ fermions. We show here how this may be done. 

As a starting point, consider the polar decomposition of the transfer matrix for the doubled Ising model, which takes the form
\begin{equation}\label{spinT2}
\hat{T}_{C}\hat{T}_{D}=\sum_{i,j=1}^{2^{M}}|L_{iC}\rangle \otimes |L_{jD}\rangle e^{(\lambda_{i}+\lambda_{j})L}
\langle R_{jD}|\otimes \langle R_{iC}|\,.
\end{equation}
Here, $\{|L_{iC}\rangle \otimes |L_{jD}\rangle\}$ and $\{\langle R_{jD}|\otimes \langle R_{iC}|\}$ are two complete, orthonormal sets of many-particles states for the $C$ and $D$ fermions, which in general are not bi-orthogonal.
The factors $e^{\lambda_i L}$ are the singular values of the transfer matrix for a single copy of the spin system, and the limiting values of $\lambda_i$ for large $L$ are the Lyapunov exponents characterising the spin system. For economy, we use the same symbol to denote both the disorder-dependent $\lambda_i$ at finite $L$ and its limiting value as $L\to \infty$. Since we are concerned with the largest few singular values, we adopt the ordering $\lambda_{1}\geq \lambda_{2}\geq \cdots \geq \lambda_{2^{M}}$.

Comparing Eq.\,(\ref{2ndQT}) with Eq.\,(\ref{spinT2}), one sees that the values taken by $\exp(\alpha^{\dagger}_{i}\alpha_{i}[\sigma^{z}\otimes \epsilon L]_{ii})$ for $\alpha^{\dagger}_{i}\alpha_{i}=0$ or $1$ generate the possible values of $e^{(\lambda_{i}+\lambda_{j})L}$. In particular, ignoring for the moment questions connected with parity, the largest of the Lyapunov exponents for the doubled Ising model is obtained by setting $\alpha^{\dagger}_{i}\alpha_{i}=1$ for $1\leq i \leq M$ and $\alpha^{\dagger}_{i}\alpha_{i}=0$ for $M+1\leq i \leq 2M$. The associated right vector is
\begin{equation}\label{Psi}
|\Psi\rangle\equiv|R_{1C}\rangle \otimes |R_{1D}\rangle=
\prod_{i=1}^{M} \frac{1}{\sqrt{2}}[W^{T}_{ij}f^{\dagger}_{j}+V_{ij}g^{\dagger}_{j}]|\mbox{vac}\rangle,
\end{equation}
where $|\mbox{vac}\rangle$ is the vacuum for $f$ and $g$ fermions, and for simplicity we have omitted the subscript $R$ on $W$ and $V$.
The state $|\Psi\rangle$ satisfies for all $i$ the equations
\begin{equation}\label{defeq1}
\left[ W_{ij}^{T}f^{\dagger}_{j}+V_{ij}g^{\dagger}_{j}\right]|\Psi\rangle=0
\end{equation}
and
\begin{equation}\label{defeq2}
\left[ W_{ij}^{T}f_{j}-V_{ij}g_{j}\right]|\Psi\rangle=0.
\end{equation}
Let $P=\frac{1}{2}[W^{T}-V]$ and $Q=\frac{1}{2}[W^{T}+V]$.
Taking the difference between Eq.\,(\ref{defeq1}) and Eq.\,(\ref{defeq2}) yields $\gamma_{i}^{\dagger}|\Psi\rangle=0$ for all $i$, where the fermion creation operators $\gamma^{\dagger}_{i}$ are defined by
\begin{equation}\label{bogo}
\gamma^{\dagger}_{i}=P_{ij}C_{j}+Q_{ij}C_{j}^{\dagger}.
\end{equation}
(Of course, similar expressions for the $D$ system may be obtained from the sum of Eq.\,(\ref{defeq1}) and Eq.\,(\ref{defeq2}).)
In this way we find that the right vector associated with the largest possible singular value of the spin transfer matrix is
\begin{equation}
|R_{1C}\rangle=\prod_{i=1}^{M}\gamma_{i}^{\dagger}|0\rangle ,
\end{equation}
where $|0\rangle$ is the vacuum for the $\gamma$-fermions.
More generally, we can obtain all the right vectors as follows. First, in the factor  $\exp(\alpha^{\dagger}_{i}\alpha_{i}[\sigma^{z}\otimes \epsilon L]_{ii})$ from Eq.\,(\ref{2ndQT}), for each $i$ in the range $1\leq i \leq M$ either: (a) set $\alpha^{\dagger}_{i}\alpha_{i}=1$ and  $\alpha^{\dagger}_{i+M}\alpha_{i+M}=0$; or (b) set $\alpha^{\dagger}_{i}\alpha_{i}=0$ and $\alpha^{\dagger}_{i+M}\alpha_{i+M}=1$.
The corresponding right vector $|R\rangle$ satisfies for (a) $\gamma^{\dagger}_i|R\rangle=0$ and for (b) $\gamma_i|R\rangle=0$.
The associated Lyapunov exponents for the (undoubled) Ising model are 
\begin{equation}\label{exponents}
\lambda_j=\sum_{i=1}^{M}\epsilon_{i}(\gamma_{i}^{\dagger}\gamma_{i}-\frac{1}{2})\,,
\end{equation}
where $\gamma_{i}^{\dagger}\gamma_{i}=1$ or $0$ for (a) and (b) respectively.

As a further step in the discussion, it is necessary to distinguish between the two sectors with even and odd parity for the fermion numbers $N_C$ and $N_D$. Except in strip geometry ($\kappa_{n,M}=0$ in Eq.\,(\ref{BT})), different boundary conditions are imposed on the network model for each sector, and so each sector has its own set of Lyapunov exponents, $\epsilon$, and matrices, $W$ and $V$. We indicate quantities calculated using boundary conditions appropriate for even and odd parity sectors with plus and minus signs respectively: $\epsilon^{\pm}$, $W^{\pm}$ and $V^{\pm}$. Introducing the number operator for $\gamma$ fermions, $N_{\gamma} = \sum_{i=1}^{M} \gamma^{\dagger}_i \gamma_i$, it is straightforward to see that, in general, either $\exp({i\pi N_c}) = \exp({i\pi N_{\gamma}})$ or $\exp({i\pi N_c}) = -\exp({i\pi N_{\gamma}})$, but to determine which of these holds in a particular instance requires explicit (numerical) calculation.
To this end, we consider (restricting ourselves for simplicity to even $M$) the scalar product of $|\Psi\rangle$ (see Eq.\,(\ref{Psi})), for which we know that $e^{i\pi N_{\gamma}}|\Psi\rangle=+|\Psi\rangle$, with a reference state, $|\mbox{ref}\rangle$, chosen in order that $\exp({i\pi N_c})|\mbox{ref}\rangle=+|\mbox{ref}\rangle$. The result $\langle\mbox{ref}|\Psi\rangle \not= 0$ will indicate $e^{i\pi N_c} = e^{i\pi N_{\gamma}}$, while (barring accidental orthogonality) the result $\langle\mbox{ref}|\Psi\rangle= 0$ implies $\exp({i\pi N_c}) = -\exp({i\pi N_{\gamma}})$. A suitable choice for $|\mbox{ref}\rangle$ is the state
\begin{equation}
|\mbox{ref}\rangle =\prod_{i}^{M}(f_i^{\dagger} + g_i^{\dagger})|\mbox{vac}\rangle\,,
\end{equation}
which satisfies $N_C |\mbox{ref}\rangle = N_D |\mbox{ref}\rangle = M |\mbox{ref}\rangle$ and hence also $\exp({i\pi N_c})|\mbox{ref}\rangle=+|\mbox{ref}\rangle$. The scalar product is
\begin{equation}
\begin{array}{rl}
\langle\mbox{ref}|\Psi\rangle&= 2^{-M/2} \mbox{det}(W^{T}+V)\vspace{0.3cm}\\
 &= 2^{-M/2}  \mbox{det}(W)\mbox{det}(1+WV)\,.
\end{array}
\end{equation}
The only factor on the right side of this expression which may be zero is $\mbox{det}(1+WV)$. It turns out that $\chi\equiv \mbox{det}(WV)$, which takes the values $\chi=\pm 1$, is a convenient indicator: barring accidental degeneracies in the spectrum of $WV$, $\mbox{det}(1+WV)=0$ if and only if  
$\chi=-1$. 

The proof of this statement is as follows. One has
\begin{equation}
\label{sgndet}
\mbox{det}(1+WV)=\prod_{i}(1+\rho_{i})
\end{equation}
where $\rho_{i}$ are the eigenvalues of the O(M) matrix $WV$. These occur as complex conjugate pairs, $\rho_i$ and $\rho_i^*$, and possibly also as real pairs, $1$ and $-1$, of which there will be at most one in the absence of degeneracy. If there {\it is} one such real pair, $\chi=-1$ and $\mbox{det}(1+WV)=0$; if there is none, $\chi=+1$ and $\mbox{det}(1+WV)\not=0$.

We now apply these results to obtain expressions for the Lyapunov exponents of the Ising model transfer matrix in terms of those of the network model. 
For simplicity of presentation we make use of a property which appears to hold generally and is certainly true for the model studied in Sec.\ref{rbim}, the $\pm J$-RBIM with $p\leq 0.5$. In this system,  $\chi^+=+1$ always, and half of the  Lyapunov exponents $\lambda_i$ are obtained from Eq.\,(\ref{exponents}) by setting $\epsilon \equiv \epsilon^+$ and taking $N_{\gamma}$ even. The remaining exponents result from setting  $\epsilon \equiv \epsilon^-$, accompanied by even $N_{\gamma}$ if $\chi^-=+1$, and by odd $N_{\gamma}$ if $\chi^-=-1$. Since we are concerned in the following only with $\chi^-$, we write it below simply as $\chi$.

Using the expression for the exponents, Eq.\,(\ref{exponents}), we find the following rules for the case $\chi=1$
\begin{equation}\label{lambdaplus}
\begin{array}{l}
 \lambda_{1}=\frac{1}{2}\sum_{i=1}^{M} \epsilon_{i}^{+} \vspace{0.2cm}\\
 \lambda_{2}=\frac{1}{2}\sum_{i=1}^{M} \epsilon_{i}^{-}\ -\ \epsilon^{-}_{1} \vspace{0.2cm}\\
 \lambda_{3}=\frac{1}{2}\sum_{i=1}^{M} \epsilon_{i}^{-}\ -\ \epsilon^{-}_{2} \vspace{0.2cm}\\
 \lambda_{4}=\frac{1}{2}\sum_{i=1}^{M} \epsilon_{i}^{+}\ -\ \epsilon^{+}_{1}\ -\ \epsilon^{+}_{2}.
\end{array}
\end{equation}
For the case $\chi=-1$, we have instead
\begin{equation}\label{lambdaminus}
\begin{array}{l}
 \lambda_{1}=\frac{1}{2}\sum_{i=1}^{M} \epsilon_{i}^{+} \vspace{0.2cm}\\
 \lambda_{2}=\frac{1}{2}\sum_{i=1}^{M} \epsilon_{i}^{-} \vspace{0.2cm}\\
 \lambda_{3}=\frac{1}{2}\sum_{i=1}^{M} \epsilon_{i}^{-}\ -\ \epsilon^{-}_{1}\ -\ \epsilon^{-}_{2}\vspace{0.2cm}\\
 \lambda_{4}=\frac{1}{2}\sum_{i=1}^{M} \epsilon_{i}^{+}\ -\ \epsilon^{+}_{1}\ -\ \epsilon^{+}_{2},
\end{array}
\end{equation}
where the order of $\lambda_{3}$ and $\lambda_{4}$ has to be decided numerically. 

\begin{figure}[ht]
\begin{center}
\begin{tabular}{c}
\epsfig{file=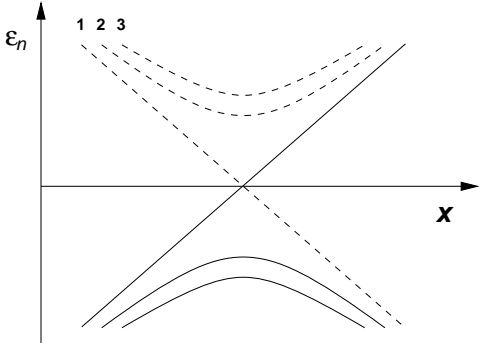,width=2.5in}
\end{tabular}
\caption{\small A sign change of $\chi$ as a function of a parameter $x$ is accompanied by the smallest Lyapunov exponent reaching zero. This
may be regarded as a form of level crossing, as illustrated. 
\label{levcross}}
\end{center}
\end{figure}

It is interesting to note a consequence that follows from the importance of $\chi$, and which is probably characteristic of localisation problems in class D. It arises if $\chi$ can change sign as a continuous parameter, such as temperature in the Ising model, is varied. Since the two subspaces of $WV\in {\rm O(M)}$ in which $\chi=+1$ and $\chi=-1$, respectively, are disconnected, a change in the sign of $\chi$ is accompanied by the vanishing of $\epsilon_{1}^{-}$. This process is a form of level crossing, as illustrated in Fig.\ref{levcross}. In the RBIM it occurs for large $M$ at the Curie point, as discussed in Sec.\,\ref{rbim}.

This distinction between phases with either sign for $\chi$ is the analogue for the RBIM in cylindrical geometry of a topological classification introduced for two-dimensional systems from class D in Ref.\onlinecite{readgreen} and for one-dimensional, single channel systems in Ref.\onlinecite{motrunich}. In particular, such one-dimensional systems may have two phases: in one phase a long sample supports a zero-energy state at each of its ends, and in the other it does not. Turning to the network model for large $L$, we note that the combinations $V_L^TW_L^T$ and $W_RV_R$ are the reflection matrices from either end of the system. A closed sample may be constructed in an obvious way, by joining outgoing links to ingoing links in pairs at each end of the system. For a network model, a stationary state has the status of a zero energy state, and stationary states will exist at the ends of the closed sample if the reflection matrices for the corresponding open system have $1$ as an eigenvalue. From the discussion following Eq.\,(\ref{sgndet}), one sees that this is the case if $\chi=-1$ but not if $\chi=+1$.

\section{Calculational methods}\label{NSC}

\subsection{Numerical procedure}\label{sec.numeth}

Numerical methods suitable for studying random transfer matrix products in general are well-established and described, for example, in Refs.\onlinecite{benettin,pichard,KrMcKi}. It has been recognised recently,\cite{chalker} however, that these methods may develop an instability to rounding errors and must be modified when applied to systems in symmetry class D. Specifically, the modifications are required if the smallest positive Lyapunov exponent approaches zero on a scale set by the spacing between other exponents, which happens in the RBIM at the Curie point, as described in the Sec.\,\ref{LE} and Sec.\,\ref{rbim}. We summarise the established algorithm and review the modification required in this subsection.

First, we define some notation. Consider a network model of width $2M$ links and length $L$, with a transfer matrix of the form given in Eq.\,(\ref{T1}). Let ${\bf x}^{k}(n)$, for $k=1,2,\cdots,2M$ and $n$ fixed, be orthonormal column vectors, each of $2M$ components, written in the same basis as this transfer matrix. These vectors are generated by a sequence of operations designed to ensure that ${\bf x}^{k}(L)$ converges for large $L$ to the $k$-th column of the matrix 
\begin{equation}\label{matrix}
\frac{1}{\sqrt 2}\left(\begin{array}{cc}
W_{L} & -W_{L} \vspace{0.3cm}\\
V^{T}_{L} & V^{T}_{L} \end{array} \right) 
\end{equation}
appearing in the polar decomposition, Eq.\,(\ref{T1}).

The conventional choice\cite{benettin,pichard,KrMcKi} for these operations is as follows. Pick ${\bf x}^{k}(0)$ arbitrarily. With $n=0$, let 
\begin{equation}
{\bf y}^{k}=T(L-n-s,L-n)\,{\bf x}^{k}(n)\,,
\label{algorithm}
\end{equation}
and perform Gram-Schmidt orthonormalisation, following
\begin{equation}\label{ortho1}
{\bf z}^{k}={\bf y}^{k}-\sum_{i=1}^{k-1}([{\bf x}^{i}(n+s)]^T\cdot
{\bf y}^{k}){\bf x}^i(n+s)
\end{equation}
and 
\begin{equation}\label{ortho2}
{\bf x}^k(n+s)={\bf z}^{k}/|{\bf z}^{k}|\,.
\end{equation}
The process is repeated with $n=s, 2s \ldots L-s$. The Lyapunov exponents are then the mean growth rates
\begin{equation}\label{LYP}
\epsilon_{k}=\langle \frac{1}{s}\ln \left|{\bf z}^{M+1-k}\right| \rangle
\equiv -\langle \frac{1}{s}\ln \left|{\bf z}^{M+k}\right| \rangle
\end{equation}
for $k=1 \ldots M$, where the average is over successive orthonormalisation steps. 
The interval $s$ is taken for computational efficiency to be as large as is possible without rounding errors significantly affecting the orthogonalisation. 

The rate of approach with increasing $L$ of the vectors ${\bf x}^i(L)$ to the columns of Eq.\,(\ref{matrix}) is determined by the spacing between successive Lyapunov exponents. So also are the deviations at large $L$ of these vectors from the columns of Eq.\,(\ref{matrix}). Such deviations are induced by numerical noise and generate errors in the calculated values of Lyapunov exponents. 
For systems in symmetry class D, the value of the smallest positive Lyapunov exponent, $\epsilon_1$, may approach zero. If it does, the vectors ${\bf x}^M(L)$ and ${\bf x}^{M+1}(L)$ are unusually susceptible to rounding errors, as is the value of $\epsilon_1$ determined from Eq.\,(\ref{LYP}). We demonstrate in Appendix \ref{round} that the error decreases with  decreasing noise amplitude, $\sigma$, only as $|\ln(\sigma)|^{-1}$. Because of this, a modification must be found that stabilises the algorithm.

Following Ref.\onlinecite{chalker}, we adapt the Gram-Schmidt orthonormalisation to enforce the $2 \times 2$ block structure evident in Eq.\,(\ref{matrix}). Denoting the $j$-th component of ${\bf x}^{k}(n)$ by ${x}_j^{k}(n)$, and similarly for ${\bf y}^k$ and ${\bf z}^k$, we replace
Eq.\,(\ref{ortho1}) for $1\leq k \leq M$ by
\begin{equation}
{z}_j^{k}={y}_j^{k}-\sum_{i=1}^{k-1}\Big(\sum_{l=1}^M{x}_l^{i}(n+s)
{y}_l^{k}\Big){x}_j^i(n+s)
\end{equation}
if $1\leq j \leq M$, and by
\begin{equation}
{z}_j^{k}={y}_j^{k}-\sum_{i=1}^{k-1}\Big(\sum_{l=M+1}^{2M}{x}_l^{i}(n+s)
{y}_l^{k}\Big){x}_j^i(n+s)
\end{equation}
if $M+1\leq j \leq 2M$. Similarly, we replace Eq.\,(\ref{ortho2}) by
\begin{equation}
{x}_j^k(n+s)={{z}_j^{k}}/{\Big[\sum_{i=1}^{M}|{z}_i^{k}|^2\Big]^{1/2}}\,,
\end{equation}
if $1\leq j \leq M$, and by
\begin{equation}
{x}_j^k(n+s)={{z}_j^{k}}/{\Big[\sum_{i=M+1}^{2M}|{z}_i^{k}|^2\Big]^{1/2}}\,,
\end{equation}
if $M+1\leq j \leq 2M$. Lyapunov exponents are determined as before from Eq.\,(\ref{LYP}), and now remain stable to rounding errors even if $\epsilon_1 \to 0$.

\subsection{Self-averaging quantities}\label{physics}

We wish to calculate for the Ising model the free energy, spin correlation functions, and correlations of disorder operators. In the presence of bond randomness these all exhibit sample-to-sample fluctuations, but the free energy density and typical decay lengths appearing in correlations functions are self-averaging. In this subsection we describe how such self-averaging quantities can be obtained from the Lyapunov exponent spectrum of the network model. The calculation of correlation functions themselves is discussed in Sec.\,\ref{correlation2}. 

We start from the polar decomposition of the transfer matrix for an (undoubled) Ising model of width $M$ and length $L$, which (in analogy to Eq.\,(\ref{spinT2})) is
\begin{equation}\label{spinT}
\hat{T}=\sum_{l}^{2^M}|L_{l}\rangle e^{\lambda_{l}L} \langle R_{l}|\,.
\end{equation}
Defining the reduced free energy per site as 
\begin{equation}
F=-\lim_{L \to \infty} \ln(Z)/LM
\end{equation}
and using Eq.\,(\ref{partition}), Eq.\,(\ref{lambdaplus}) and Eq.\,(\ref{lambdaminus}), we have by standard arguments\begin{equation}\label{freee}
F=-\lim _{L\to\infty} [ \frac{1}{LM}\ln {\cal |A|} +\frac{1}{2M}\sum_{i=1}^{M}\epsilon_{i}^{+}]\,.
\end{equation}  

Turning our attention to typical decay lengths, we note first that, viewing the network model as a localisation problem, its smallest positive Lyapunov exponent defines a localisation length $\xi$ through $\xi\equiv \epsilon_1^{-1}$. In a localised phase $\xi$ has a finite limit, the bulk localisation length, as $M\to \infty$, while at a mobility edge one expects that $\xi$ diverges with $M$ and that a universal scaling amplitude, $a$, is defined by the limiting value of $M\epsilon_1$ for $M\to \infty$. An unusual feature of localisation problems in symmetry class D is that one may have $a=0$; from the discussion of Sec.\,\ref{LE} and results presented in Sec.\,\ref{rbim}, this occurs in the RBIM in the sector with odd parity.

For the Ising model, the typical correlation length $\xi_{\sigma \sigma}$ appearing in the spin-spin correlation function may be extracted as follows. This correlator, for two spins with (in the 
notation of Fig.\,\ref{ising}) the same coordinates in the 
vertical direction and separation $n-m$ in the horizontal direction, is
\begin{equation}\label{spincorreln}
%\begin{array}{l}
\langle \sigma^{x}_{i}(0)\sigma^{x}_{i}(n)\rangle= 
\frac{\mbox{Tr} [\sigma_i^{x}\hat{T}(1,n)\sigma_i^{x}\hat{T}(n+1,L)]}
{\mbox{Tr}[\hat{T}(1,L)]}\,.
%\end{array}
\end{equation}
Recalling that $\sigma^{x}_{i}$ has non-zero matrix elements 
only between states with opposite parity, and taking $L \to \infty$,  $\xi_{\sigma\sigma}$ is defined and expressed in terms of the 
Lyapunov exponents for the spin transfer matrix by
\begin{equation}\label{ssd}
\xi_{\sigma\sigma}^{-1}=-\lim_{n\to\infty}\frac{1}{n}
\ln(\langle \sigma^{x}_{i}(0)\sigma^{x}_{i}(n)\rangle)
= \lambda_{1}-\lambda_{2}\,.
\end{equation}
When writing $\xi_{\sigma\sigma}$ in terms of the network model 
Lyapunov exponents, it is useful to introduce a lengthscale 
$\xi_{\rm 1D}$ which characterises the sensitivity of the network model to changes in boundary conditions, and is defined by
\begin{equation}
\xi_{\rm 1D}^{-1}=\frac{1}{2}\sum_{i}[\epsilon_{i}^{+}-\epsilon_{i}^{-}]\,.
\end{equation}
We expect insensitivity to boundary conditions except at the 
critical point, and anticipate that $\xi_{\rm 1D}^{-1}\sim \exp(-M/\xi)$
for large $M$.
In regions of the RBIM phase diagram for which $\chi=1$ 
(corresponding, as we argue, to the paramagnet), we have from Eq.\,(\ref{lambdaplus})
\begin{equation}
\xi_{\sigma\sigma}^{-1}=\epsilon^{-}_{1}+\xi_{1D}^{-1}\,,
\end{equation}
so that asymptotically the localisation length, $\xi$, and spin correlation length, $\xi_{\sigma \sigma}$, 
are equal. By contrast, in regions of the phase diagram for which $\chi=-1$ (corresponding to the ferromagnet) we have $\xi_{\sigma\sigma}=\xi_{\rm 1D}$. This large lengthscale here characterises the decay of spin 
correlations in a quasi-one dimensional sample within the ordered phase 
of the two-dimensional system. Such decay is governed by rare domain-wall excitations that cross the width of the sample. Because $\xi_{\sigma\sigma}$ 
is large when $\chi=-1$, it is useful also to examine the inverse lengthscale governing corrections to Eq.\,(\ref{ssd}), which is $\lambda_1-\lambda_3$. 
For $\chi=-1$
\begin{equation}
\lambda_{1}-\lambda_{3}=\epsilon^{-}_{1}+\epsilon^{-}_{2}+\xi_{\rm 1D}^{-1}\,,
\end{equation}
so that, as $M\to \infty$, $\epsilon^{-}_{1}+\epsilon^{-}_{2}$ gives 
the typical decay rate of the connected part of the spin 
correlation function in 
the ordered phase.

In a similar way, one can obtain $\xi_{\mu \mu}$, the typical correlation length for the disorder operators $\mu_r$ of Kadanoff and Ceva.\cite{kadce} These operators are defined at points $r$ which lie at the centres of plaquettes in the Ising model. The two-point correlation function $\langle \mu_0 \mu_r \rangle$ is defined by considering a modified system in which exchange interactions crossed by a path on the dual lattice between $0$ and $r$ have their sign changed. Then
\begin{equation}\label{disorder}
\langle \mu_0 \mu_r \rangle = \frac{Z'}{Z}
\end{equation}
is the ratio of the partition function $Z'$ for the modified system to that of the original system, and
\begin{equation}\label{xi-mu-mu}
\xi_{\mu \mu}^{-1}= -\lim_{r \to \infty}\frac{1}{r}\ln(\langle \mu_0 \mu_r \rangle)\,.
\end{equation}
Because the different boundary conditions imposed on the network model in sectors of even and odd parity constitute an (infinite) line of such modified bonds, $\xi_{\mu \mu}$ may be expressed in terms of $\epsilon^+$ and $\epsilon^-$. Moreover, in the ferromagnetic phase ($\chi=-1$), $\xi_{\mu \mu}$ is the reduced interfacial tension between domains of opposite magnetisation. To make this explicit, let $F_{p}$ and $F_{a}$ be reduced free energies per site, calculated from the definition Eq.\,(\ref{freee}) for systems in cylindrical geometry with, respectively, periodic ($\sigma_{M+1}^x\equiv \sigma_{1}^x$) and antiperiodic ($\sigma_{M+1}^x\equiv -\sigma_{1}^x$) boundary conditions on spins imposed around the cylinder. Then
\begin{equation}\label{interface}
\xi_{\mu \mu}^{-1}=M(F_p-F_a)\,.
\end{equation}
In this phase, we find using the ideas of Sec.\ref{LE} that
\begin{equation}
\xi_{\mu \mu}^{-1} = \epsilon^{-}_{1}+\xi_{1D}^{-1}\,,
\end{equation}
while in the paramagnetic phase ($\chi=+1$)  we obtain $\xi_{\mu \mu} = \xi_{\rm 1D}$, so the decay length diverges with $M$. As one might expect, the behaviour of $\xi_{\mu \mu}$ in each phase is similar to that of $\xi_{\sigma \sigma}$ in the dual phase.

\subsection{Correlation functions}\label{correlation2}

Calculation of the full form of correlation functions is more involved than that of the typical decay lengths since, of course, the results cannot be expressed solely in terms of Lyapunov exponents. Nevertheless, it turns out that even powers of correlation functions may be determined straightforwardly.\cite{readlud} In the most important example of the second power, one requires the product of two equivalent correlation functions, evaluated for each of the two copies of the Ising model that are combined in the network model. In the case of the square of the two-point correlation function of disorder operators, Eq.\,(\ref{disorder}), this means that the same modification of bonds is introduced in both copies of the Ising model, so that $(Z')^2$ is determined from a network model with a specific set of modified nodes. In the case of the square of the spin-spin correlation function, one can take a similar route by expressing this in terms of a disorder correlator in a dual system. Alternatively, one can write the product of two copies of a spin operator in terms of $f$ and $g$ fermions, as we describe below. By either route, one arrives ultimately at the same result: the square of the spin-spin correlation function is given by the ratio of the square of a partition function calculated from a modified network model to same quantity calculated from an unmodified model. By contrast, odd powers of correlation functions, including the first power, appear to be much harder to evaluate, leaving the sign of the correlation function undetermined: we summarise the difficulties that arise at the end of this subsection.  

\begin{figure}[ht]
\begin{center}
\begin{tabular}{ccc}
\epsfig{file=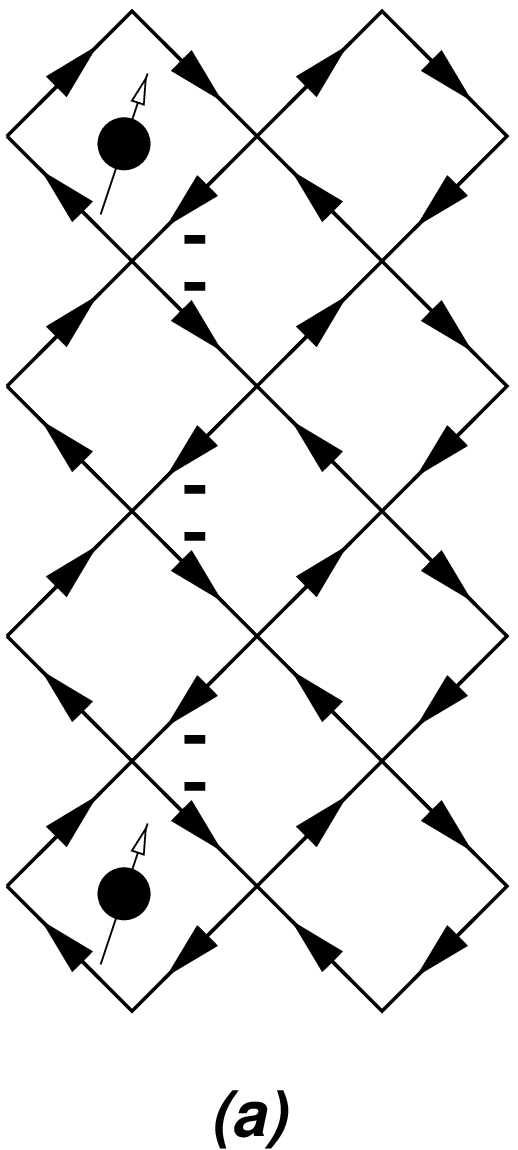,width=1.0in} 
$\quad$ &
\epsfig{file=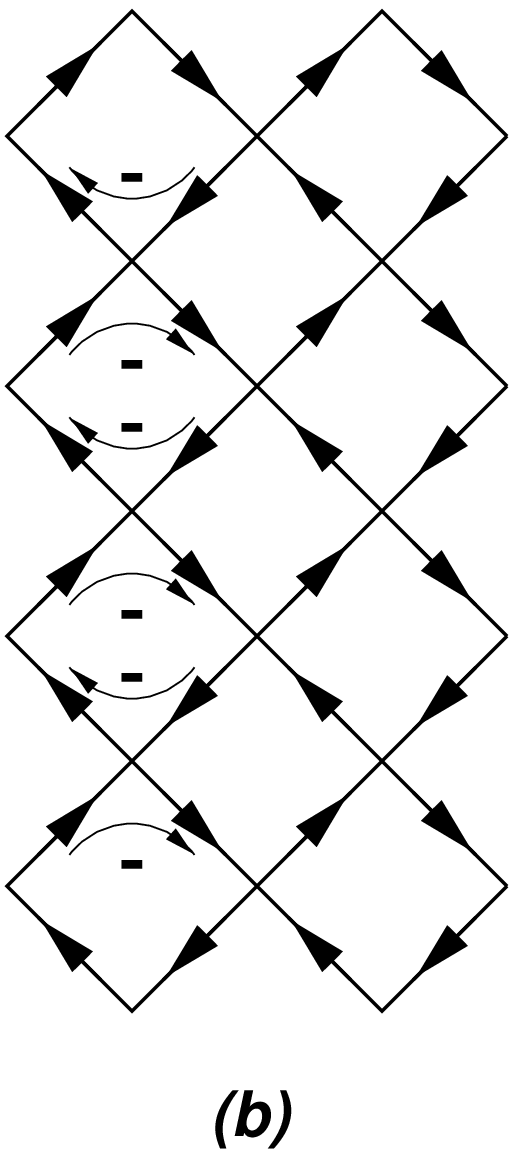,width=1.0in}
$\quad$ &
\end{tabular}
\parbox{6.5in}{\caption{Schematic representation of the effect of modifying the system by inserting spin operators in one slice: (a) link phases associated with a pair of spin operators; (b) equivalent node phases. \label{plaq}}}
\end{center}
\end{figure}

To obtain the squared spin-spin correlation function
%$\langle \sigma_i^x(n) \sigma_j^x(m) \rangle^2$ 
following Eq.\,(\ref{spincorreln}), we must evaluate products 
involving the transfer matrix for the doubled Ising model and also factors of the form 
$\sigma_{jC}^x\sigma_{jD}^x$. 
From Eqs.\,(\ref{JW}) and (\ref{summtr}) we have 
\begin{equation}
\sigma_{jC}^{x}\sigma_{jD}^{x}
= {i}(-1)^{j} 
\exp\Big(i\pi\ \sum_{i=1}^{j-1} [f^{\dagger}_{i}g_{i}+g^{\dagger}_{i}f_{i}]+i\pi f^{\dagger}_{j}f_{j}\Big)\,.
\end{equation}
In the spirit of Sec.\ref{ntwk} we translate this into first-quantised form. Each operator $\exp(i\pi[f^{\dagger}_{i}g_{i}+g^{\dagger}_{i}f_{i}])$ is represented by a $2\times 2$ matrix, $\exp(i\pi \sigma^x)=-\openone$. As a result, on one slice of the network model phase factors of $-1$ are associated with each of the right and left going links having coordinate $i$ in the range $1\leq i \leq j-1$. In addition, the operator $\exp(i\pi f^{\dagger}_jf_j)$ is represented by a similar phase factor associated with the $j$-th right-going link. These phase factors are illustrated schematically in Fig.\,(\ref{plaq}a), using as an example the combination $\sigma_{1C}^x \sigma_{1D}^x \sigma_{4C}^x \sigma_{4D}^x$, which arises in the calculation of $\langle \sigma_i^x(n) \sigma_j^x(m) \rangle^2$ on setting $n=m$, $i=1$ and $j=4$. Such link phases can equally be attributed to nodes representing vertical bonds of the Ising model, as indicated in Fig.\,(\ref{plaq}b). Viewed in this way, the insertion of spin operators into the transfer matrix product is represented by a change in node parameters $\kappa_{n,i} \rightarrow \kappa_{n,i} + i\pi/2$ for $1\leq i \leq j-1$. In turn, this is equivalent to a change in sign for the corresponding dual bond strengths, as it should be since the spin correlation function can be evaluated as a disorder correlator in the dual model.

Implementing this approach in numerical calculations, we determine the singular values of the transfer matrix for modified and unmodified network models of length $L$, of course using the same realisation of disorder for both. From these we calculate the largest singular value of the transfer matrix for the doubled spin system, which we denote by $\exp(2\lambda_1'L)$ in the modified case, and by $\exp(2\lambda_1L)$ in the unmodified case, following the notation of Sec.\,\ref{LE}. For large $L$
\begin{equation}
\langle \sigma_i^x(n) \sigma_j^x(m) \rangle^2 = \exp(2[\lambda_1'-\lambda_1]L)\,.
\end{equation}
In practice, the combination $[\lambda_1' - \lambda_1]L$ approaches a finite limiting value rather quickly with increasing $L$. Conveniently, it is not necessary to evaluate the scalar products of the form $|\langle L_1|R_1\rangle|^2$ which appear in the numerator and denominator of Eq.\,(\ref{spincorreln}), because for large $L$ these are the same in the modified and unmodified systems, and therefore cancel.

As mentioned above, calculation of the unsquared spin-spin correlation function presents greater practical problems. A route is clear in principle: one can use the discussion of Sec.\ref{LE} to construct the transfer matrix for the undoubled Ising model, via its polar decomposition, in terms of Slater determinants of the $\gamma$ fermions; and one can also express spin operators in this Ising model in terms of the creation and annihilation operators for these fermions. Difficulties then arise from the fact that the matrices $W_L$ and $W_R$ appearing in Eq.\,(\ref{PD}) are unrelated in the presence of disorder, as also are $V_L$ and $V_R$. In consequence, one has to deal with two sets of $\gamma$ fermions: $\gamma_L$ and $\gamma_R$. Put briefly, we find (as in Eq.\,(4.7) of Schultz, Mattis and Lieb\cite{SML}) that  $\langle \sigma_i^x(n) \sigma_j^x(n) \rangle$ can be written as an expectation value of a product of $2|i-j|$ fermion operators, which can be evaluated using Wick's theorem. However, in the disordered system it is not possible to reduce this expectation value to a single determinant (as in Eq.\,(4.13) of Ref\onlinecite{SML}). Without such a reduction, the computational effort required to determine $\langle \sigma_i^x(n) \sigma_j^x(n)\rangle$ seems prohibitive for large $|i-j|$.

\section{Numerical results for the $\pm J$-RBIM}\label{rbim}

\subsection{Introduction}

In this section we present results obtained using the mapping from the Ising model to the network model as a way of studying the $\pm J$ RBIM. Previous work of this type has been described by Cho\cite{chothesis}, but without the advantages of the numerical algorithm or the detailed relation between the network model and statistical mechanical quantities that we have discussed in Sec.\,\ref{NSC}. The $\pm J$ RBIM, defined on a square lattice, has nearest-neighbour exchange couplings $J_{ij}$ 
drawn independently from the probability distribution
\begin{equation}
P(J_{ij})=(1-p)\delta(J_{ij}-J) + p\delta(J_{ij}+J)\,,
\end{equation}
with $0\leq p \leq 1$ and $J$ positive; we set $J=1$ in the following.

\begin{figure}[ht]
\begin{center}
\begin{tabular}{c}
\epsfig{file=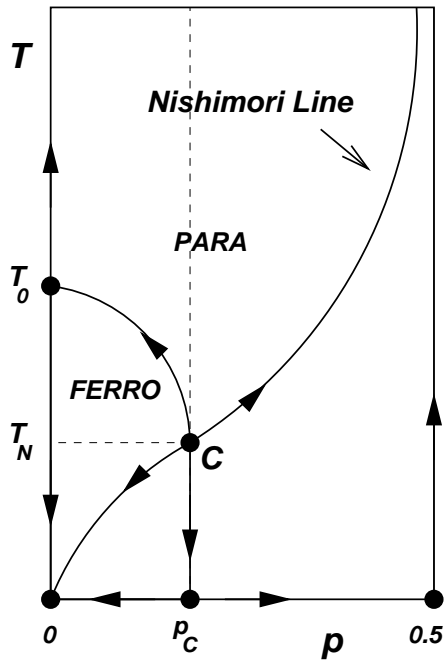,width=2.3in}
\end{tabular}
\parbox{3.3in}{\caption{Phase diagram of the $\pm J$-RBIM with superimposed RG scaling flow.\label{flow}}}
\end{center}
\end{figure}

The phase diagram of the model, as a function of temperature $T$ and the concentration $p$ of antiferromagnetic bonds, is shown in  Fig.\,\ref{flow}, with renormalisation group (RG) scaling flow superimposed.\cite{youngsouthern,rammal,nishiorg,nishi,LeDouss,kitavert,singh,OzNishi1,houdayer,morgbin,mcmillan,OzNishi,UenoOz,KitaOgu,queiroz,honecker,Ozeki,KawaRieger}
The pure system ($p=0$) has a transition
between ferromagnetic and paramagnetic phases at a Curie temperature $T_{0}=2[$ln$(1+\sqrt{2})]^{-1}$. As antiferromagnetic bonds are introduced the Curie temperature is depressed, and the ferromagnetic phase is destroyed altogether above a threshold concentration $p_{c}$. A curve in the $p-T$ plane known as the Nishimori line\cite{nishiorg,nishi,LeDouss,kitavert} (NL) plays an important role in the discussion of scaling flow. It is defined for the $\pm J$ RBIM by the equation $\exp(2\beta J)=(1-p)/p$. On this line the RBIM has an additional gauge symmetry, because of which the internal energy is analytic 
and ensemble-averaged spin-spin correlations obey the equalities $[\langle\sigma_i^x(n)\sigma_j^x(m)\rangle^{2k-1}]=[\langle\sigma_i^x(n)\sigma_j^x(m)\rangle^{2k}]$ for integer $k$. The NL cuts the phase boundary separating the ferromagnet from the paramagnet at a point $C$, the Nishimori point, with coordinates $p_{c},T_{N}$. This point is particularly interesting as an example of a disorder-dominated multicritial point. One of the two scaling flow axes in its vicinity lies along the NL, while the other coincides with the phase boundary,\cite{LeDouss} as indicated in Fig.\,\ref{flow}. Scaling flow on the critical manifold for $p<p_c$ runs from the Nishimori point towards the critical fixed point of the pure system, at which disorder is marginally irrelevant. The phase boundary on the other side of the Nishimori point is believed to be vertical\cite{nishi,LeDouss,kitavert} in the $T-p$ plane, and on it the scaling flow runs from the Nishimori point towards a zero-temperature critical point. Finally, the phase diagram for $p>1/2$ can be obtained from that shown for $p<1/2$ by reflection in the line $p=1/2$, using a gauge transformation which maps $p$ to $1-p$ and the ferromagnetically ordered phase to an antiferromagnet.
 
Despite the considerable effort which has been invested in studies of the RBIM, some aspsects of its behaviour are not yet well-characterised. In the following, we present a high-accuracy determination of the position of the phase boundary and of critical properties at the Nishimori point.

\subsection{Method}

We use the numerical method described in Sec.\,\ref{sec.numeth} to calculate the Lyapunov exponents of the network model associated with the RBIM, studying two copies of the system for each disorder realisation, with boundary conditions appropriate for fermion numbers of each parity. In the spirit of Sec.\,\ref{physics} we use the smallest positive exponent calculated for the network model with periodic boundary conditions to define a characteristic inverse lengthscale, and analyse the finite-size scaling behaviour of $M/\xi=M\epsilon^{-}_{1}$ as a function of system width $M$. In addition, we determine the interfacial tension, $\xi_{\mu \mu}^{-1}$, and study its size-dependence. We also calculate the disorder-averaged square of the spin-spin correlation function, $[\langle \sigma^x_i(n) \sigma^x_j(n)\rangle^2]$, for spins lying in the same slice of the system, using the approach described in Sec.\,\ref{correlation2}.

For most of the results presented, we study system widths in the range from $M=8$ to $M=256$ spins, and system lengths of $L=5\times 10^5$ spins. Realisation-dependent fluctuations in self-averaging quantities decrease as $L^{-1/2}$ and in some cases increase with $M$. As an example, using $L=5\times 10^5$ the value of $\epsilon^{-}_{1}$ at the Nishimori point is obtained with an accuracy of $1\%$ for $M=16$ and $2\%$ for $M=64$. Some calculations require higher precision. In particular, the high-resolution studies of the interfacial tension close to the Nishimori point, presented in Sec.\,\ref{nishi-collapse}, and of scaling on the phase boundary, presented in Sec.\,\ref{pb-collapse}, use systems of length up to $L=2\times 10^8$, restricting accessible widths to $M\leq 32$.

\subsection{Location of phase boundary}\label{phase-boundary}

In this subsection we describe the determination of the form of the boundary between the ferromagnetic and paramagnetic phases. We also discuss the nature of finite-size effects in different parts of the phase diagram. For this purpose the quantity $\chi$, introduced in Sec.\ref{LE}, is very useful and we substantiate our claim that (in the thermodynamic limit) the sign of $\chi$ indicates which phase the Ising model is in. 

\begin{figure}[ht]
\begin{center}
\begin{tabular}{c}
\epsfig{file=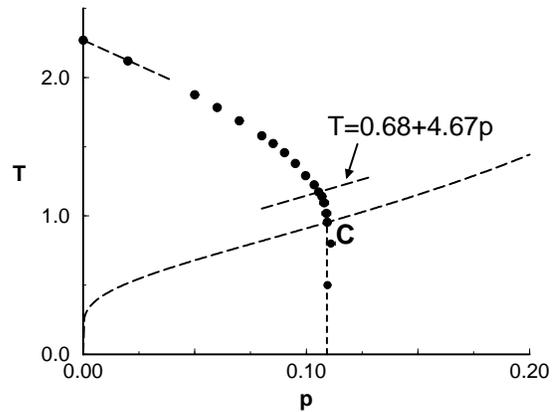,width=3.1in}
\end{tabular}
\parbox{3.3in}{\caption{The location of the phase boundary determined from  numerical calculations. Data obtained on the line $T=0.68+4.67p$ are presented in Fig.\,\ref{12a} 
\label{diagram}}}
\end{center}
\end{figure}
\begin{figure}[ht]
\begin{center}
\begin{tabular}{c}
\epsfig{file=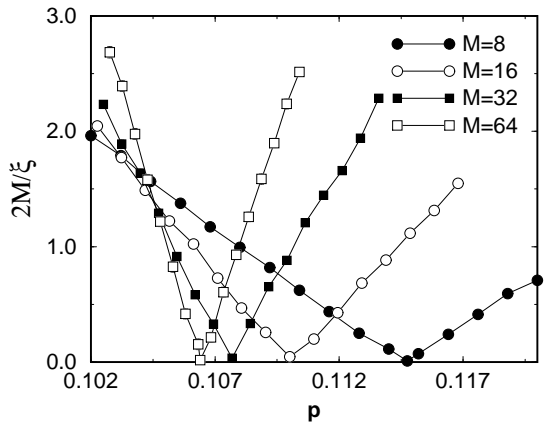,width=3.1in}
\end{tabular}
\parbox{3.3in}{\caption{Values of $2M/\xi$ calculated crossing the phase boundary along the line $T=0.68+4.67p$.\label{12a}}}
\end{center}
\end{figure}

Our results for the position of the phase boundary are shown in Fig.\ref{diagram} and in Table \ref{list}. Points on this phase boundary are found from a finite-size scaling analysis of the variation of $M/\xi$ along lines that intersect it; the slopes of these lines in the $p-T$ plane are chosen to avoid crossing the boundary at small angles. Representative data, calculated on the line $T=0.68+4.67p$, are shown in Fig.\ref{12a}; they have two features that can be used to identify the boundary. First, the curves of $M/\xi$ for two successive values of $M$ have an intersection point, and with increasing $M$ these intersection points approach the boundary from the small-$p$ side. 
Second, for each $M$, there is a value of $p$ at which $\xi$ diverges, or equivalently $\epsilon^-_1=0$. With increasing $M$, these points approach the boundary from the large-$p$ side. We obtain consistent results using the two methods.

A test of these calculations follows from the fact that the tangent to the ferromagnetic-paramagnetic boundary at the pure critical point is
known exactly\cite{domany} to be $dT_{p}/dp|_{p=0}=-7.2821...$. From a linear approximation at $p=0.005$ we find 
$dT_{p}/dp|_{p=0}=-7.32\pm0.06$, in good agreement with this. Our values for $T_p$ are also compatible with those given in Ref.\onlinecite{queiroz}.

\renewcommand{\arraystretch}{1.2}
\begin{table}[ht]
\begin{center}
\begin{tabular}{ll|ll}
%\begin{tabular}{@{\hspace{0.7cm}}l@{\hspace{1.2cm}}l|@{\hspace{0.3cm}}l@{\hspace{0.9cm}}l}
$p$ & $T_{p}$ & $p$ & $T_{p}$\\ \hline
0.005 & 2.2325 $\pm$ 0.0003 & 0.0903 $\pm$ 0.0002 & 1.458 \\
0.02 & 2.120 $\pm$ 0.001 & 0.0951 $\pm$ 0.0005 & 1.379 \\
0.05 & 1.875 $\pm$ 0.001 & 0.1000 $\pm$ 0.0005 & 1.294 \\
0.06 & 1.783 $\pm$ 0.002 & 0.1035 $\pm$ 0.0011 & 1.224 \\
0.07 & 1.688 $\pm$ 0.002 & 0.1055 $\pm$ 0.0011 & 1.173 \\
0.08 & 1.580 $\pm$ 0.002 & 0.1080 $\pm$ 0.0021 & 1.095 \\
0.0852 & 1.523 $\pm$ 0.002 & 0.1090 $\pm$ 0.0021 & 1.019 \hspace{0.3cm}\\ 
\end{tabular} 
\end{center}
\parbox{3.3in}{\caption{Location of the phase boundary.\label{list}}}
\end{table}

\begin{figure}[ht]
\begin{center}
\begin{tabular}{c}
\epsfig{file=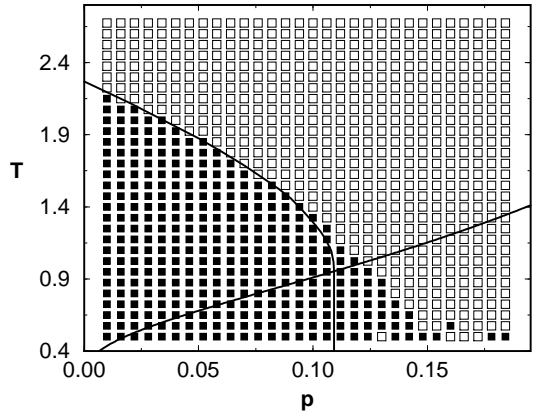,width=3.1in}
\end{tabular}
\parbox{3.3in}{\caption{The sign of $\chi$ for
a system of width $M=16$ as a function of position in the $T-p$ plane. Open squares indicate $\chi=+1$ and filled squares $\chi=-1$. The NL and phase boundary are also shown.\label{chidiagram}}}
\end{center}
\end{figure}

It is evident from the data shown in Fig.\ref{12a}, and its equivalent for other values of $p$ and $T$, that $\epsilon_1^-=0$ along a line in the phase diagram which approaches the phase boundary for large $M$, but is displaced from it into the paramagnetic phase for finite $M$. From the discussion given in Sec.\ref{LE}, we expect $\chi$ to change sign on this same line, being for large $M$ positive in the paramagnetic phase and negative in the ferromagnetic phase. 
The data shown in Fig.\ref{chidiagram} demonstrates that this is so; Fig.\ref{chidiagram} also shows that the finite-size shift in the position of the phase boundary is
very large in the portion of the phase diagram lying below the NL. It seems possible that these finite-size effects may provide an alternative explanation of data which have been interpreted\cite{OzNishi,UenoOz} as evidence for a random antiphase state\cite{rammal} lying in this region of the phase diagram; and it seems likely that they are responsible for non-monotonic temperature-dependence of Lyapunov exponents for the RBIM, reported at $p>p_c$ in Ref.\onlinecite{queiroz}.
 
\subsection{Nishimori Line}\label{nishi-collapse}

In this subsection we examine critical behaviour near the multicritical point $C$ as it is approached along the Nishimori line. The facts\cite{nishi,LeDouss} that $C$ is known to lie on the NL, and that the NL coincides with one of the scaling flow axes at $C$, both greatly help the analysis. We obtain consistent, high-accuracy estimates of the coordinate $p_c$ and the exponent $\nu$ using three separate analyses of the finite-size scaling of $\xi$ and also from a study of the interfacial tension.

\begin{figure}[ht]
\begin{center}
\begin{tabular}{c}
\epsfig{file=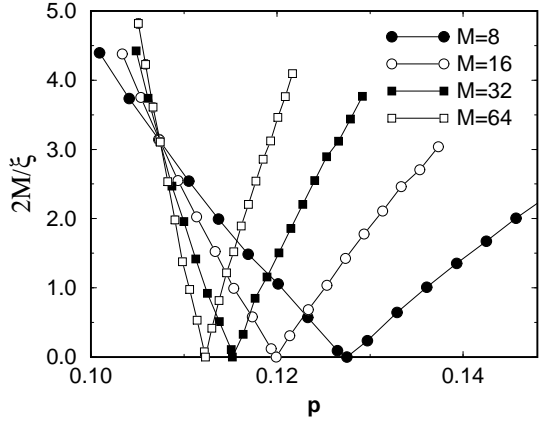,width=3.1in}
\end{tabular}
\parbox{3.3in}{\caption{Variation of $2M/\xi$ along the NL, with position parameterised by $p$.\label{NL}}}
\end{center}
\end{figure}
An overview of the variation of $M/\xi$ along the NL is given in Fig.\ref{NL}. We apply finite-size scaling ideas to the data in the following different ways.
Two of them are similar to the methods used in Sec.\ref{phase-boundary} to locate the phase boundary: first, curves of $M/\xi$ for two successive values of $M$ cross, and we focus on these crossing points for increasing $M$; second, for each $M$ there is a point on the NL at which $M/\xi=0$, and we study the position of these points as a function of $M$. Third, we can collapse data for different $M$ and from the whole critical region onto a single curve. 

\begin{figure}[ht]
\begin{center}
\begin{tabular}{c}
\epsfig{file=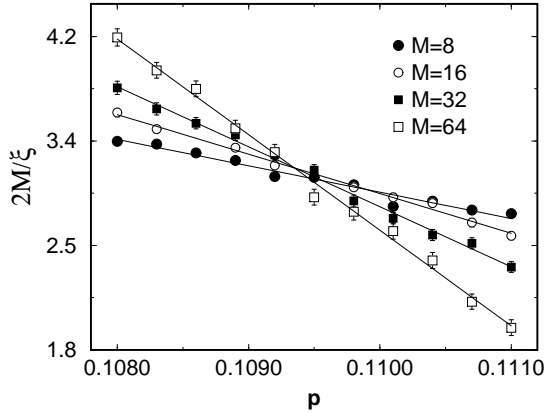,width=3.1in}
\end{tabular}
\parbox{3.3in}{\caption{Variation of $2M/\xi$ along the NL, close to the Nishimori point.\label{intersec}}}
\end{center}
\end{figure}
Turning to the first of these, we concentrate on the top left of Fig.\ref{NL}, where data sets intersect roughly at one point. Behaviour in this region is shown on a larger scale in Fig.\ref{intersec}. From an extrapolation of the intersection points to large $M$ we find $p_{c}=0.1093\pm0.0002$.
We also obtain a limiting value at the intersection point of  $M/\xi=1.58\pm0.01$ as $M \to \infty$. The value of $\nu$ may be found from the scaling with $M$ of the gradients of curves at the intersection points; a similar analysis can also be made for the interfacial tension and we present both together, towards the end of this subsection.

\begin{figure}[ht]
\begin{center}
\begin{tabular}{c}
\epsfig{file=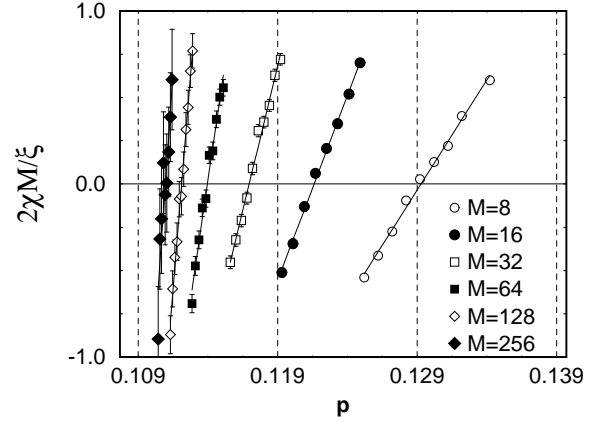,width=3.1in}
\end{tabular}
\parbox{3.3in}{\caption{Variation of the combination $2\chi M/\xi$ along the NL.\label{extratop}}}
\end{center}
\end{figure}
\begin{figure}[ht]
\begin{center}
\begin{tabular}{c}
\epsfig{file=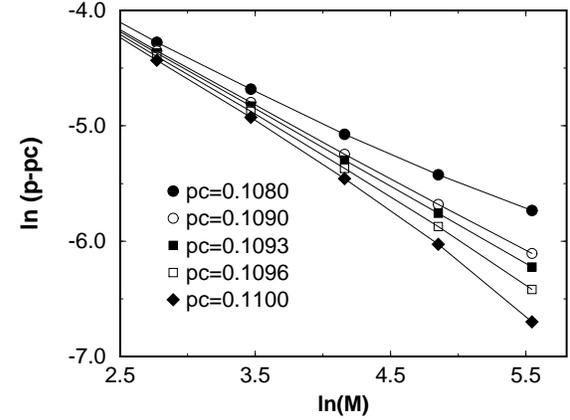,width=3.1in} 
\end{tabular}
\parbox{3.3in}{\caption{$\ln(p_{M}-p_c)$ as a function of $\ln(M)$, for different estimates of $p_c$. The straightest line is obtained with $p_{c}=0.1093$ and has slope $-1.49$.
\label{extrabot}}}
\end{center}
\end{figure}

Taking a second approach to the data, the points $p_M$ on the NL at which $M/\xi=0$ are determined for $8\leq M \leq 256$ as shown in Fig.\ref{extratop}, where we take advantage of the fact that, for fixed $M$, the combination $\chi M/\xi$ varies smoothly through zero as a function of position along the NL. One expects the finite-size shift $p_{M}-p_c$ to vary with $M$ as $(p_{M}-p_c)\propto M^{-1/\nu}$, and we show
the dependence of $p_{M}-p_c$ on $M$ in Fig.\ref{extrabot}, using a double logarithmic scale for various choices of $p_{c}$. With the correct choice for $p_c$, this data should fall onto a straight line of slope $-\nu$. By this method we find $p_{c}=0.1093$ and $\nu=1.49\pm 0.05$.

\begin{figure}[ht]
\begin{center}
\begin{tabular}{c}
\epsfig{file=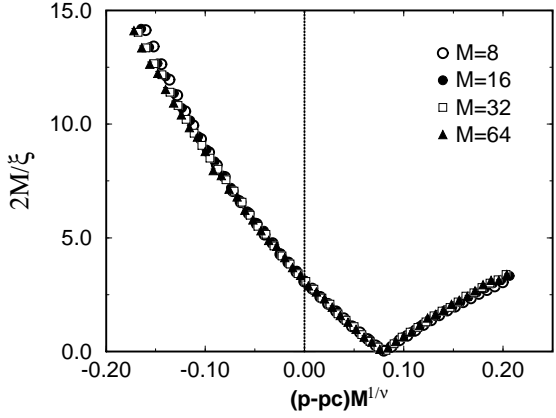,width=3.1in}
\end{tabular}
\parbox{3.3in}{\caption{Data collapse along the NL, using $\nu= 1.50$ and $p_{c}=0.1093$.\label{colla}}}
\end{center}
\end{figure}

\begin{figure}[ht]
\begin{center}
\begin{tabular}{c}
\epsfig{file=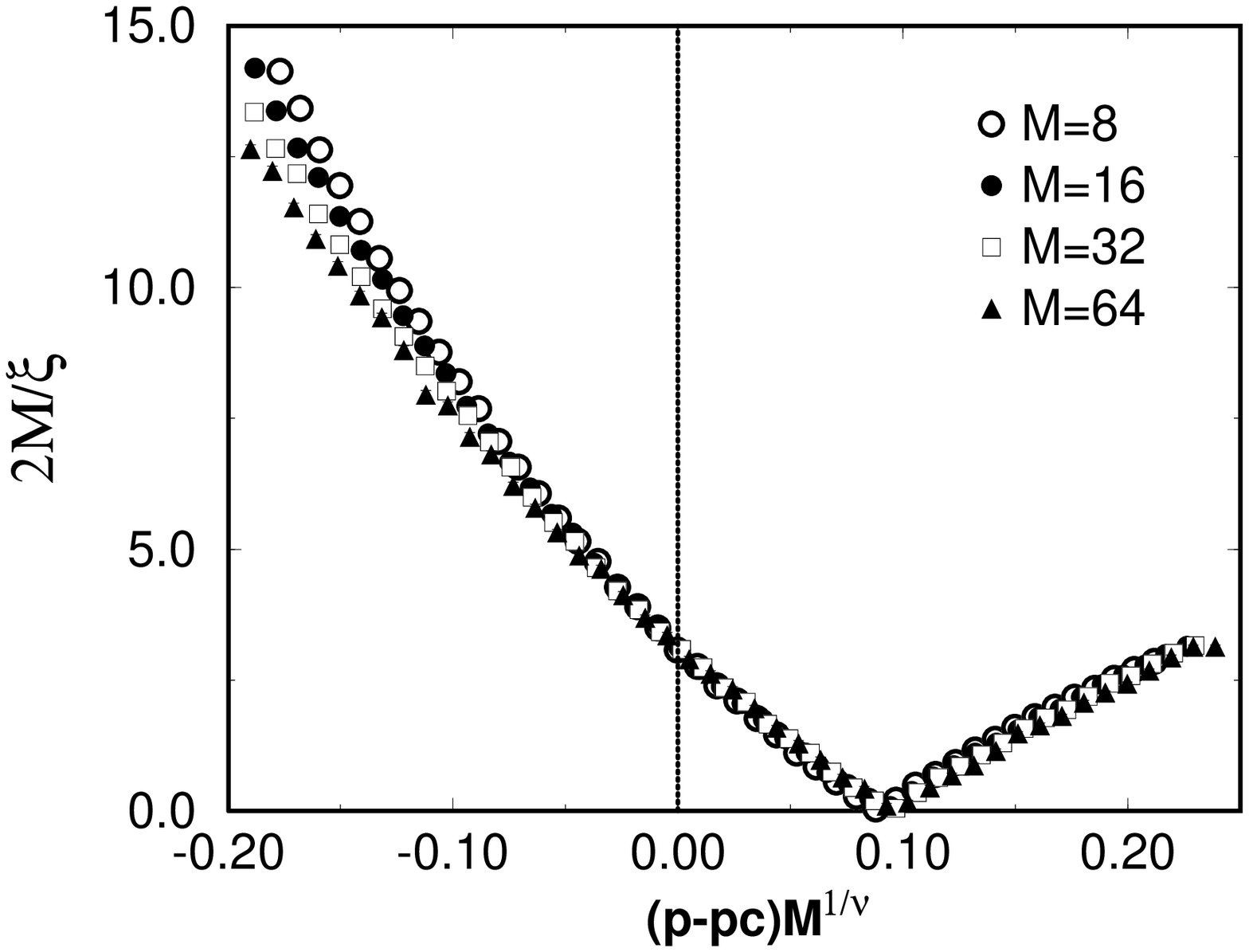,width=3.1in}
\end{tabular}
\parbox{3.3in}{\caption{Data collapse along the NL, using $\nu= 1.40$ and $p=0.1093$.\label{colla140}}}
\end{center}
\end{figure}

A third treatment of the data for $M/\xi$ is provided by attempting to collapse all points from the critical region of Fig.\,\ref{NL} onto a single curve, plotting $M/\xi$ as a function of $(p-p_c)M^{1/\nu}$. In principle, both $p_c$ and $\nu$ may be taken as fitting parameters, but we find that $p_c$ is more accurately determined using the methods described earlier. We therefore set  $p_{c}=0.1093$ and vary only the value of $\nu$. We find the best collapse, shown in Fig.\ref{colla}, taking $\nu=1.50$. Visibly worse collapse results from using $\nu=1.40$, as shown in Fig.\ref{colla140}; by such comparisons we find $\nu=1.50\pm 0.10$, confirming the result derived from Fig.\ref{extrabot}, but not improving on it in accuracy. 

\begin{figure}[ht]
\begin{center}
\begin{tabular}{c}
\epsfig{file=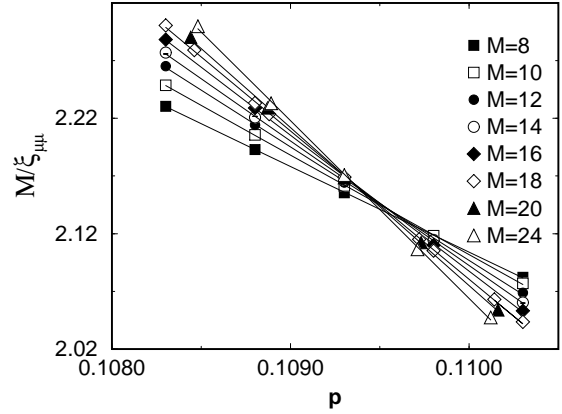,width=3.1in}
\end{tabular}
\parbox{3.3in}{\caption{Variation of $M/\xi_{\mu \mu}$ on the NL close to the Nishimori point.\label{rawdat}}}
\end{center}
\end{figure}

\begin{figure}[ht]
\begin{center}
\begin{tabular}{c}
\epsfig{file=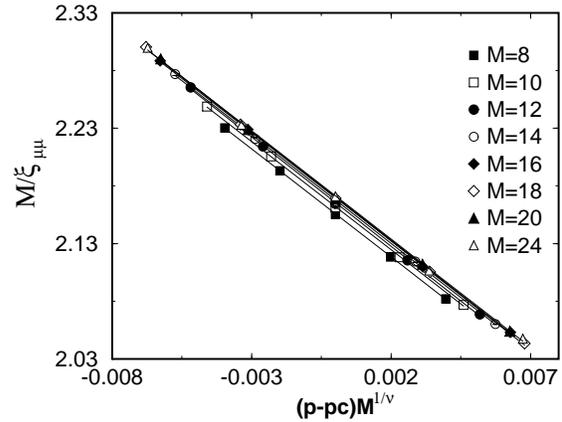,width=3.1in}
\end{tabular}
\parbox{3.3in}{\caption{Scaling of $M/\xi_{\mu \mu}$  as a function of $(p-p_c)M^{1/\nu}$, using $\nu=1.50$ and $p_{c}=0.1093$.\label{honeck-scaling}}}
\end{center}
\end{figure}

\begin{figure}[ht]
\begin{center}
\begin{tabular}{c}
\epsfig{file=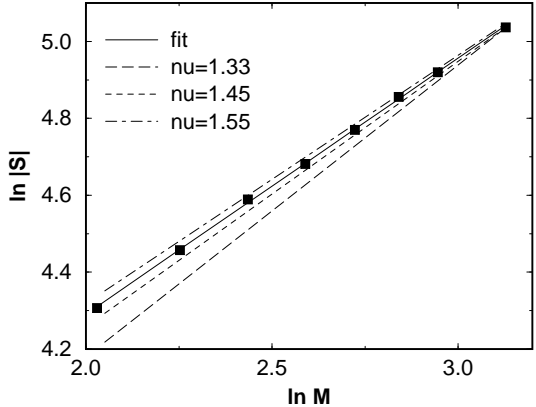,width=3.1in}
\end{tabular}
\parbox{3.3in}{\caption{Scaling of the gradient $S$ of lines in Fig.\ref{rawdat} as a function of $M$. The best fit inverse slope is $\nu=1.50$
(solid line). Lines corresponding to $\nu=1.33,1.45$ and $1.55$ are also shown. 
\label{slopes}}}
\end{center}
\end{figure}

Finally as a way to check the conclusions we have reached from finite-size scaling of $M/\xi$, and in order to make a direct comparison with recent work by Honecker, Picco and Pujol,\cite{honecker} we present a study of the interfacial tension, $\xi^{-1}_{\mu \mu}$, defined in Eq.\,(\ref{xi-mu-mu}). High-precision data, calculated using $L=2\times 10^8$ for $8\leq M \leq 24$ on the NL very close to the Nishimori point, are shown in Fig.\ref{rawdat}; statistical errors are smaller than symbol sizes.
As with $M/\xi$, one expects, in the critical region and at sufficiently large $M$, to collapse data for $M/\xi_{\mu \mu}$  onto a single curve by plotting it as a function of the scaling variable $(p-p_c)M^{1/\nu}$. Such a collapse is illustrated in Fig.\ref{honeck-scaling}, using $p_{c}=0.1093$ and $\nu=1.50$.
Deviations from collapse are evident at smaller values of $M$, appearing as vertical offsets of the corresponding lines in Fig.\ref{honeck-scaling}. Corrections to scaling of this type are expected, and arise from scaling variables which are irrelevant in the RG sense at the critical point: in general, we have
\begin{equation}\label{scalingcorr}
M^2\Delta f = a + b(p-p_c)M^{1/\nu} + cM^{-x} +\ldots
\end{equation}
where $x$ is the exponent associated with the leading irrelevant scaling variable, $a$ is a universal scaling amplitude, and $b$ and $c$ are constants.
Such corrections occur at the pure Ising transition,\cite{Sorensen} and have also been studied in the U(1) network model.\cite{eastmondhuckestein} In view of the way that they enter Eq.\,(\ref{scalingcorr}), it is appropriate to concentrate on the $M$-dependence of the gradients of lines in Fig.\,\ref{rawdat} when determining $\nu$. These gradients are shown as a function $M$ using a double logarithmic scale in Fig.\ref{slopes}, from which we derive our most precise estimate of $\nu$, $\nu=1.50 \pm 0.03$.

\begin{figure}[ht]
\begin{center}
\begin{tabular}{c}
\epsfig{file=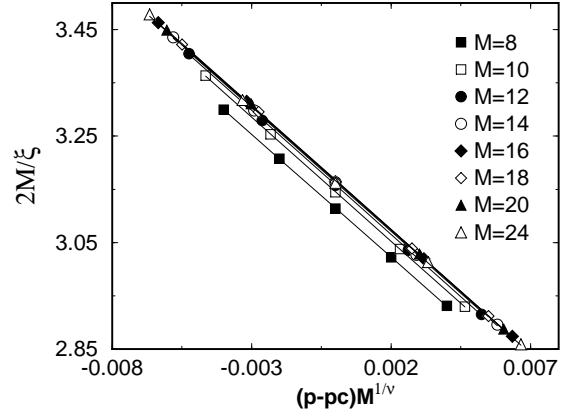,width=3.1in}
\end{tabular}
\parbox{3.3in}{\caption{Scaling of $M/\xi$ on the  NL close the the Nishimori point, using $\nu=1.50$ and $p_{c}=0.1093$\label{epsilon-scaling}}}
\end{center}
\end{figure}

The scaling of $M/\xi$ close to the critical point can be analysed in just the same way, yielding the same result for $\nu$. This scaling collapse is depicted in Fig.\,\ref{epsilon-scaling}.

We conclude our analysis of critical behaviour on the Nishimori line with the results: $p_{c}=0.1093\pm 0.0002$ and $\nu=1.50\pm 0.03$. 
Our value for $p_c$ is consistent with the result $p_{c}=0.1094\pm0.0002$,
obtained by Honecker, Picco and Pujol,\cite{honecker}
who carried out a detailed study of the interfacial tension and correlation functions, using the Ising model transfer matrix in a spin basis, which restricted system widths to $M\leq 12$.
Our value for $p_c$ is also in agreement with some earlier, less precise values, including
$p_{c}=0.111\pm 0.002$, in Ref.\onlinecite{OzNishi} and $p_c=0.1095\pm0.0005$ in Ref.\onlinecite{queiroz}, both found using a transfer matrix approach with up to 14 spins. It is also 
marginally in agreement with $p_{c}=0.104$ from Ref.\onlinecite{KawaRieger} obtained as the critical disorder strength around $T=0$.
It is in marginal disagreement with the result from series expansions,\cite{singh} $p_c=0.114 \pm 0.003$. 
More strikingly, however, our value for $\nu$ is in disagreement with previous estimates, which lie close\cite{singh} to the percolation value,  $\nu=4/3$, including most recently $\nu=1.33\pm0.03$ in Ref.\onlinecite{honecker}.
We believe that the larger system sizes accessible in our work, and the allowance we have made for irrelevant scaling variables at the critical point, together account for the discrepancy, and that the data shown in Figs.\,\ref{colla140} and \ref{slopes} exclude this smaller value of $\nu$.

\subsection{Scaling along the phase boundary}\label{pb-collapse}

The phase boundary separating the ferromagnet from the paramagnet coincides\cite{LeDouss} with the second relevant scaling axis at the Nishimori point, in addition to that defined by the NL. On the boundary, we expect scaling flow from $C$ towards
the pure critical point for $p<p_c$, and from $C$ towards the zero-temperature critical point for $T<T_N$. We analyse such flow in this subsection.

\begin{figure}[ht]
\begin{center}
\begin{tabular}{c}
\epsfig{file=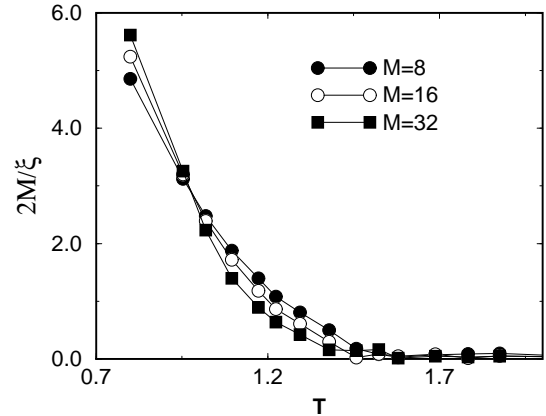,width=3.1in}
\end{tabular}
\parbox{3.3in}{\caption{Variation of $M/\xi$ with position, parameterised by $T$, on the phase boundary. \label{crossover}}}
\end{center}
\end{figure}

Qualitative evidence in support of these established ideas is presented in 
Fig.\ref{crossover}, which shows the variation of $M/\xi$ with position, parameterised by $T$, on the phase boundary, and with $M$. For $p<p_c$, the coordinates of points on the phase boundary are taken from Table \ref{list}, while for $T<T_N$ we assume the phase boundary to be vertical in the $p-T$ plane and set $p=p_c$, using our estimate for the value of $p_c$. At temperatures $T>T_N \simeq 0.9533$, $M/\xi$ decreases with increasing $M$, approaching zero which is the value taken by this scaling amplitude in the pure Ising model at $T=T_0\simeq 2.269$; fluctuations visible in Fig.\ref{crossover} for data at temperatures $T\gtrsim 1.5$ arise from errors in determining the position of the phase boundary.
At the Nishimori point itself, curves of $M/\xi$ for different $M$ cross, with a limiting value for $M\to \infty$, as already determined in our study of behaviour on the NL. For $T<T_N$, values of $M/\xi$ increase both with decreasing $T$ and with increasing $M$, as expected if flow is towards lower temperatures.

\begin{figure}[ht]
\begin{center}
\begin{tabular}{c}
\epsfig{file=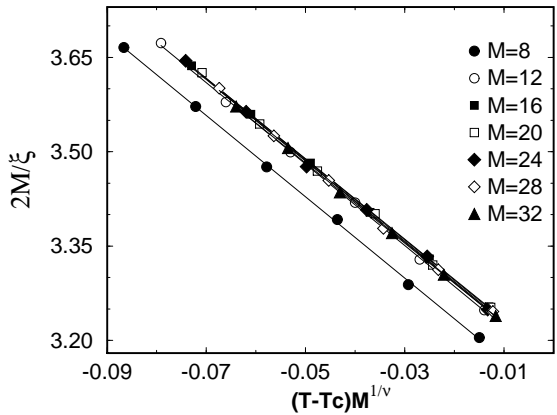,width=3.1in}
\end{tabular}
\parbox{3.3in}{\caption{Scaling of $M/\xi$ on the phase boundary below the Nishimori point, using $\nu_{T}=4.0$ and $T_{N}=0.9533$\label{other}}}
\end{center}
\end{figure}

Scaling flow along the phase boundary close to the Nishimori point is characterised by a critical exponent $\nu_T$, which in principle can be determined using an approach similar to that taken for $\nu$. In practice, there are extra difficulties. First, in contrast to the NL, the form of the phase boundary is not known exactly; we choose the simpler regime, $T<T_N$, and set $p$ to our estimate for $p_c$, as above. Second, it happens that $\nu_T>\nu$, so that flow away from the multicritical point is faster in the direction of the NL than along the phase boundary. Because of this, the range for $T$ over which useful data can be collected is limited on both sides. The distance, $T_N-T$, from the Nishimori point should not be too large, or data will lie outside the critical region. It should not be too small, either, because close to $C$ errors in our value for $p_c$ will be dominant. Having limited the range for $T-T_N$ in this way, the variation in $M/\xi$ is also restricted. It is therefore particularly important that statistical errors are small, and so we study samples of length $L=2\times 10^8$ with $8\leq M \leq 32$. The scaled data are presented in Fig.\,\ref{other}: as with the analysis presented in Fig.\,\ref{honeck-scaling} and Fig.\,\ref{epsilon-scaling}, and as expected from Eq.\,\ref{scalingcorr}, the value of $\nu_T$ is determined mainly from the gradients of curves for each $M$. We conclude that $\nu_{T}=4.0\pm 0.5$. While this confidence margin is wide, it is encouraging that on extrapolating the data in Fig.\ref{other} to $T=T_N$ we obtain at the Nishimori point $M/\xi = 1.58\pm 0.01$ for $M\to \infty$, in perfect agreement with the value found independently from data collapse on the NL.

\subsection{Behaviour at strong disorder}

In three or more dimensions, the random bond Ising model has a spin-glass phase at low temperature and strong disorder.\cite{binderyoung} It is known that spin-glass order does not occur in the two-dimensional RBIM, except at zero temperature,\cite{binderyoung} but it is of interest to examine behaviour at strong disorder using the methods we have developed. 

\begin{figure}[ht]
\begin{center}
\begin{tabular}{c}
\epsfig{file=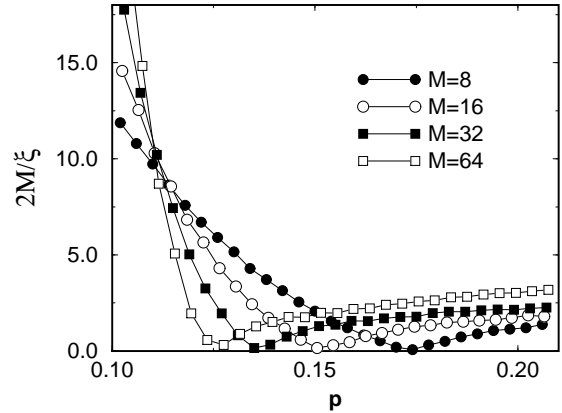,width=3.1in}
\end{tabular}
\parbox{3.3in}{\caption{Variation of $M/\xi$ with $p$, crossing the phase boundary at $T=0.5$.\label{teq}}}
\end{center}
\end{figure}
Finite-size effects in the RBIM are large at strong disorder and low temperature, as remarked in connection with Fig.\ref{chidiagram}, and as is clear from Fig.\ref{teq}, which shows the variation of $M/\xi$ with $p$ and $M$ at a fixed temperature, $T=0.5$, below the Nishimori point. Despite these finite-size effects, it is straightforward to identify the position of the phase boundary from Fig.\ref{teq}. Moreover, the size-dependence of $M/\xi$ in the paramagnetic phase at $T=0.5$ and higher temperatures is consistent with a finite limiting value for $\xi$ as $M\to \infty$, as required from the fact that the RBIM does not have a metallic phase.\cite{readlud}
\begin{figure}[ht]
\begin{center}
\begin{tabular}{c}
\epsfig{file=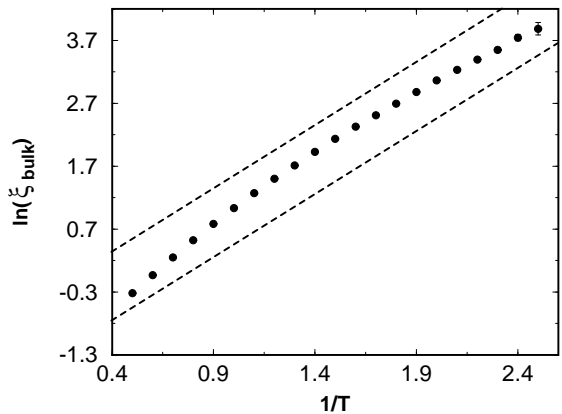,width=3.1in}
\end{tabular}
\parbox{3.3in}{\caption{Variation of $\xi_{\rm bulk}$ with $T$ on the line $p=0.5$: $\ln(\xi_{\rm bulk})$ as a function of $1/T$. Dashed lines represent $\xi_{\rm bulk} \propto \exp(-2/T)$.
\label{peq1}}}
\end{center}
\end{figure}

\begin{figure}[ht]
\begin{center}
\begin{tabular}{c}
\epsfig{file=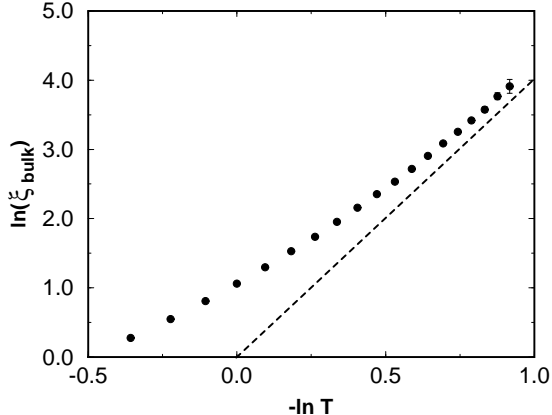,width=3.1in}
\end{tabular}
\parbox{3.3in}{\caption{Variation of $\xi_{\rm bulk}$ with $T$ on the line $p=0.5$: $\ln(\xi_{\rm bulk})$ as a function of $\ln(T)$. The dashed line represents $\xi_{\rm bulk} \propto T^{-\nu}$ with $\nu=4$.
\label{peq2}}}
\end{center}
\end{figure}

For a quantitative analysis of behaviour in this region, we focus on the line $p=0.5$ which, by symmetry arguments, is an exact scaling axis. Scaling flow is from the zero-temperature fixed point at $p=0.5$ towards infinite temperature, and one can collapse data on this line to extract the limiting behaviour of $\xi$ for $M\to\infty$. This extrapolated localisation length, $\xi_{\rm bulk}$,
is expected to be finite for $T>0$. Its temperature dependence for $T\geq 0.4$ (obtained using $8\leq M \leq 64$ and $L=10^6$) is shown in Fig.\,\ref{peq1}, where we compare our results with the behaviour $\xi_{\rm bulk} \propto \exp(-2/T)$, suggested\cite{saulkardar,bray} for the $\pm J$ RBIM. In Fig.\,\ref{peq2} we compare our same results with the power-law divergence, $\xi_{\rm bulk} \propto T^{-\nu}$, expected in a RBIM with a distribution of bond strengths continuous at $J=0$, for which exponent values in the range $\nu = 3.4$ to $\nu=4.2$ have been reported previously.\cite{mcmillan,braymoore,husemorg} Our data in the temperature range accessible do not provide firm grounds to prefer one form for the temperature dependence over the other.

\subsection{Spin-spin correlations}\label{spin-correln}

\begin{figure}[ht]
\begin{center}
\begin{tabular}{c}
\epsfig{file=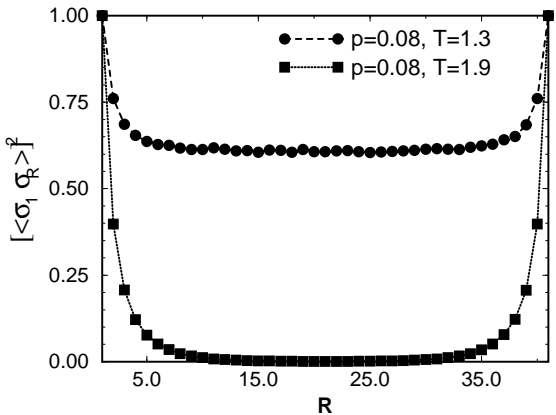,width=3.1in}
\end{tabular}
\parbox{3.3in}{\caption{Variation of the disorder-averaged square spin-spin correlation function with distance around a system of circumference $M=40$, in the paramagnetic phase ($T=1.9$) and the ferromagnetic phase $(T=1.3$).\label{corr40}}}
\end{center}
\end{figure}

As a demonstration of the effectiveness of the method set out in Sec.\,\ref{correlation2} for obtaining even powers of spin-spin correlation functions, we have calculated $[\langle \sigma_i^x(n) \sigma_j^x(n) \rangle^2]$ at all separations $|i-j|$ of spins across the width of a long system with $M=40$. Data at $p=0.08$, obtained by averaging over $10^4$ disorder realisations, are shown in Fig.\,\ref{corr40}, for a high temperature, $T=1.9$, lying in the paramagnetic phase, and for a lower temperature, $T=1.3$, lying in the ferromagnetic phase.
It is clear for this second case that the value of the square of the magnetisation can be obtained from the correlation function at separations close to $M/2$. 

We have also used this approach to calculate $[\langle \sigma_i^x(n) \sigma_j^x(n) \rangle^2]$ and $[\langle \sigma_i^x(n) \sigma_j^x(n) \rangle^4]$ on the NL at our estimated position for the Nishimori point. At this point, one expects decay of the the disorder-average of the $k$-th power of the spin-spin correlation function to be characterised by an exponent $\eta_k$. Following the analysis described in Ref.\onlinecite{honecker}, and taking $M=20$, $L$ large and $10^4$ realisations, we obtain $\eta_{2}=0.183\pm 0.003$ and $\eta_{4}=0.253\pm 0.003$, in agreement with earlier results.\cite{honecker}

\section{Summary}

To summarise, we have described in detail a mapping between the two-dimensional random-bond Ising model and a network model with the symmetries of class D localisation problems. Building on Refs.\,\onlinecite{chofisher,chothesis,readlud,GRL} we have shown in particular how separate boundary conditions arise in the network model for sectors of the Ising model transfer matrix with even and odd parity under spin reversal, and how statistical-mechanical quantities, including the free energy per site and correlation functions, may be obtained from calculations using the network model. Amongst other things, this makes clear the sense in which the Ising model correlation length may be equated with the network model localisation length. From a computational viewpoint, calculations based on the network model are much more efficient than their equivalent using an Ising model transfer matrix in a spin basis. This is illustrated by the fact that such calculations have in the past mainly been restricted to systems of width $M\leq 14$ spins, while we present results in this paper for $M\leq 256$ spins. Applying these ideas to study the Nishimori point for the $\pm J$ RBIM, we obtain a value for the exponent $\nu$ which is significantly different from previous estimates based on much smaller systems sizes; our value excludes the possibility of a simple connection between behaviour at this critical point and classical percolation, conjectured previously.\cite{singh}
Beyond computational advantages, the equivalence between the RBIM and the network model has theoretical interest. It links the transition between paramagnet and ferromagnet to a version of the quantum Hall plateau transition, as our results illustrate. Moreover, even in quasi-one dimensional systems for which there is no sharp Curie transition, a topological distinction emerges within the network model between two separate localised phases.

\section*{Acknowledgments}

We thank N. Read for collaboration in the early stages of this work and for discussions throughout. We also thank T. Davis for his data from transfer matrix calculations in the spin basis, which provided a valuable comparison with our results. We are grateful to I. A. Gruzberg and to M. Picco for helpful correspondence. This work was supported in part by EPSRC under Grant GR/J78327.

\begin{appendix}
\section{Effect of rounding errors on Lyapunov exponents}\label{round}

The numerical results presented in this paper were obtained using a modified version of the standard algorithm for studying random matrix products, as we describe Sec.\ref{sec.numeth}. The need for such a modification stems from the instability of the standard algorithm to rounding errors if the value of the smallest positive Lyapunov exponent, $\epsilon_1$, approaches zero. The instability is extreme and it is of interest to understand how it arises. In this appendix we illustrate its origin by examining a simple model problem. 

It is sufficient to consider only products of $2 \times 2$ matrices, because the instability involves only the space spanned by the vectors associated with the pair of Lyapunov exponents smallest in magnitude, denoted by ${\bf x}^M(L)$ and ${\bf x}^{M+1}(L)$ in Sec.\ref{sec.numeth}. We therefore consider a product of random matrices, each of the form
\begin{equation}
T_{n}=\left(
\begin{array}{cc}
\mbox{cosh}\ \theta_{n} & \mbox{sinh}\ \theta_{n} \vspace{0.2cm}\\
\mbox{sinh}\ \theta_{n} & \mbox{cosh}\ \theta_{n}
\end{array}
\right)\,
\end{equation}
and drawn independently from a distribution which has $\langle \theta_n \rangle =0$ in order that the Lyapunov exponents of the matrix product are zero. To model the operation of the standard algorithm, we consider evolution of a two-component vector ${\bf v}_n$ under an analogue of Eqs.(\ref{algorithm})-(\ref{ortho2}): 
\begin{equation}\label{iter}
{\bf v}_{n+1}=\frac{T_{n}{\bf v}_{n}}{|T_{n}{\bf v}_{n}|}\,.
\end{equation}
In the absence of rounding errors, ${\bf v}_n$ converges with increasing $n$ to one of the eigenvectors of $T_n$, and so it is natural to expand ${\bf v}_n$ in this basis, writing
\begin{equation}
{\bf v}_{n}=\frac{1}{\sqrt{2}}\left(\begin{array}{c} 1 \\ 1 \end{array}\right)\cos \phi_{n}+
\frac{1}{\sqrt{2}}\left(\begin{array}{c} 1 \\ -1 \end{array}\right)\sin \phi_{n}\,.
\end{equation}
In this notation, Eq.(\ref{iter}) may be written  $\tan\phi_{n+1}=\exp(-2\theta_{n})\tan\phi_{n}$, and has fixed points $\phi_n=m\pi/2$ with $m$ integer. We concentrate on the vicinity of one of these, considering the range $0\leq \phi_n \ll 1$. Then 
$\phi_{n+1}\approx \exp(-2\theta_{n})\phi_{n}$.
We take the effect of rounding errors into account by substituting
for this the evolution equation
\begin{equation}
\phi_{n+1}= \exp(-2\theta_{n})\phi_{n}+\eta_{n}\,,
\end{equation}
where $\eta_{n}$ is random with $\langle \eta_n \rangle =0$ and $\langle \eta_n \eta_m \rangle = \delta_{mn} \sigma^2$. 

A simple treatment of the stochastic process defined in this way is sufficient for our purposes. To find approximately the limiting distribution $P(\phi_n)$ at large $n$, we divide the range under consideration for $\phi_n$ into the regimes $0\leq \phi_n < \sigma$ and  $\sigma < \phi_n$. In the former the noise dominates, generating an approximately uniform distribution for $\phi_n$. We take
\begin{equation}
P(\phi_n)=C_1\,,
\end{equation}
where $C_1$ is a constant. In the latter regime we neglect the noise and use in place of $\phi_n$ the variable $y_n=\log(\phi_n)$, taking its evolution to be
\begin{equation}\label{yn}
y_{n+1}=y_{n}-2\theta_{n}\,.
\end{equation}
Since we have chosen $\langle \theta_n \rangle =0$, this generates a uniform distribution for $y_n$ in the range $\sigma \leq y_n \leq Y$, where the upper limit $Y\sim \log(\pi/4)$ represents the point at which the linearisation of $\tan(\phi_n)$ fails, and also the boundary separating the vicinities of the fixed points of Eq.(\ref{iter}) at $\phi_n=0$ and at $\phi_n=\pi/2$. 
On transforming
back to $\phi_n$ we obtain within our approximations
\begin{equation}
\begin{array}{ccc}
P(\phi) = &
\left\{
\begin{array}{c}
C_{1} \vspace{0.2cm}\\
C_{2}/\phi
\end{array}
\right.\qquad
&
\begin{array}{l}
\mbox{for\,\,\,} 0< \phi<\sigma \vspace{0.2cm}\\
\mbox{for\,\,\,} \sigma < \phi < \pi/4
\end{array}
\end{array}
\end{equation}
where $C_{2}=C_{1}\sigma$ for continuity. $C_{1}$ is determined by the normalisation condition
\begin{equation}
\int_{0}^{\pi/4}P(\phi_n)d\phi_n=\frac{1}{2}
\end{equation}
since we may take the full range for $\phi_n$ to be $0<\phi_n<\pi/2$. We find for $\sigma\ll 1$
\begin{equation}
C_{1}\simeq [2\sigma\ln({\pi}/{4\sigma})]^{-1}\,.
\end{equation}

Now consider the effect that noise-induced departures of $\phi_n$ from the fixed point at $\phi_n=0$ have on the estimate of the Lyapunov exponent, $\epsilon$. Using $\epsilon= \langle \ln|T_{n}\mathbf{v}_{n}|\rangle$, we have 
\begin{equation}
\begin{array}{rl}
\epsilon= & \frac{1}{2}\langle\ln\left[\exp({2\theta_{n}})\cos^{2}\phi_{n}+\exp({-2\theta_{n}})\sin^{2}\phi_{n}\right]\rangle\,.
\end{array}
\end{equation}
Taking for simplicity $\theta_{n}$ and $\phi_n$ small, we find
\begin{equation}
\epsilon\simeq 4\langle\phi_n^{2}\rangle \langle\theta_n^{2}\rangle\,.
\end{equation}
In the absence of noise, $\phi_n=0$ and hence $\epsilon=0$. With noise present we must evaluate
\begin{equation}
\langle \phi_n^{2} \rangle = \int_{0}^{\pi/4} P(\phi_n) \phi_n^{2} d\phi_n\,.
\end{equation}
Using our approximate form for $P(\phi_n)$ we find, for $\sigma \ll 1$, $\langle \phi_n^2 \rangle \propto |\ln(\sigma)|^{-1}$ and hence
\begin{equation}
\epsilon\propto {|\ln(\sigma)|}^{-1}\,.
\end{equation}
Thus small rounding errors may be responsible for a large error in the value obtained for the Lyapunov exponent. In the language of this appendix, the modified algorithm described in Sec.\ref{sec.numeth} uses the known symmetry of the tranfer matrix to fix $\phi_n=0$.
\end{appendix}


\begin{thebibliography}{1}
\bibitem{baxter}
R. J. Baxter, \textit{Exactly Solved Models in Statistical Mechanics} (Academic Press, New York, 1982).
\bibitem{mccoywu}
B. M. McCoy and T. T. Wu, \textit{The two-dimensional Ising model} (Harvard University Press, Cambridge, 1973).
\bibitem{SML}
T. D. Schultz, D. Mattis and E. H. Lieb, Rev. Mod. Phys \textbf{36}, 856 (1964).
\bibitem{chofisher}
S. Cho and M. P. A. Fisher, Phys. Rev. B \textbf{55}, 1025 (1997).
\bibitem{chothesis}
S Cho, PhD thesis, UC Santa Barbara, unpublished (1997).
\bibitem{readlud}
N. Read and A. W. W. Ludwig,  Phys. Rev. B {\bf 63} 024404 (2001).
 \bibitem{GRL}
I. A. Gruzberg, N. Read and A. W. W. Ludwig, Phys. Rev. B {\bf 63} 104422 (2001).
\bibitem{dotdot}
Vik. S. Dotsenko and Vl. S. Dotsenko, Adv. Phys. \textbf{32}, 129 (1983).
\bibitem{shalaev}
B. N. Shalaev, Sov. Phys. Sol. St. {\bf 26}, 1811 (1983); Phys. Rep. {\bf 237}, 129 (1994).
\bibitem{shankar}
R. Shankar, Phys. Rev. Lett. \textbf{58}, 2466 (1987).
\bibitem{ludwig}
A. W. W. Ludwig, Phys. Rev. Lett. {\bf 61}, 2388 (1988).
\bibitem{youngsouthern}
A. P. Young and B. W. Southern, J. Phys. C {\bf 10}, 2179 (1977).
\bibitem{rammal}
R. Maynard and R. Rammal, J. Phys. Lett. (Paris) {\bf 43}, L347 (1982).
\bibitem{nishiorg}
H. Nishimori, Prog. Theor. Phys. \textbf{66}, 1169 (1981).
\bibitem{nishi} 
H. Nishimori, J. Phys. Soc. Jpn. \textbf{55}, 3305 (1986).
\bibitem{LeDouss}
P. Le Doussal and A. B. Harris Phys. Rev. Lett. \textbf{61}, 625 (1988).
\bibitem{kitavert}
H. Kitatani, J. Phys. Soc. Jpn. \textbf{61}, 4049 (1992).
\bibitem{singh}
R. R. P. Singh and J. Adler, Phys. Rev. B \textbf{54}, 364 (1996).
\bibitem{OzNishi1}
Y. Ozeki and H. Nishimori, J. Phys. Soc. Jpn. \textbf{56}, 1568 (1987).
\bibitem{houdayer}
J. Houdayer, cond-mat/0101116 (2001).
\bibitem{morgbin}
I. Morgenstern and K. Binder, Phys. Rev. B {\bf 22}, 288 (1980).
\bibitem{mcmillan}
W. L. McMillan, Phys. Rev. B {\bf 29}, 4026 (1984).
\bibitem{OzNishi}
Y. Ozeki and H. Nishimori, J. Phys. Soc. Jpn. \textbf{56}, 3265 (1987).
\bibitem{UenoOz}
Y. Ueno and Y. Ozeki, J. Stat. Phys. \textbf{64}, 227 (1991).
\bibitem{KitaOgu}
H. Kitatani and T. Oguchi, J. Phys. Soc. Jpn \textbf{61}, 1598 (1992).
\bibitem{queiroz}
F. D. A. A. Reis, S. L. A. de Queiroz and R. R. dos Santos, Phys. Rev. B \textbf{60}, 6740 (1999).
\bibitem{honecker}
A. Honecker, M. Picco and P. Pujol, cond-mat/0010143 (2000).
\bibitem{Ozeki}
Y. Ozeki, J. Phys. Soc. Jpn. \textbf{59}, 3531 (1990).
\bibitem{KawaRieger}
N. Kawashima and H. Rieger, Europhys. Lett. {\bf 39}, 85 (1997).
\bibitem{blackman}
J. A. Blackman, Phys. Rev. B \textbf{26}, 4987 (1982).
\bibitem{blackmanplus}
J. A. Blackman and J. Poulter, Phys. Rev. B {\bf 44}, 4374 (1991); 
J. A. Blackman, J. R. Goncalves, and J. Poulter, Phys. Rev. B {\bf 58}, 1502 (1998).
\bibitem{saulkardar}
L. Saul and M. Kardar, Phys. Rev. E {\bf 48}, R3221 (1993); Nucl. Phys. B {\bf 432}, 641 (1994).
\bibitem{inoue}
M. Inoue, J. Phys. Soc. Jpn. {\bf 64}, 3699 (1995).
\bibitem{Sorensen}
E. S. Sorensen, cond-mat/0006233 (2000).
\bibitem{chalcodd}
J. T. Chalker and P. D. Coddington, J. Phys. C \textbf{21}, 2665 (1988).
\bibitem{AltZirn}
M. R. Zirnbauer, J. Math. Phys. \textbf{37}, 4986 (1996);
A. Altland and M. R. Zirnbauer, Phys. Rev. B \textbf{55}, 1142 (1997).
\bibitem{bundschu}
R. Bundschuh, C. Cassanello, D. Serban, and M. R. Zirnbauer, Phys. Rev. B {\bf 59}, 4382 (1999).
\bibitem{Senthil}
T. Senthil and M. P. A. Fisher, Phys. Rev. B \textbf{61}, 9690 (2000).
\bibitem{readgreen}
N. Read and D. Green, Phys. Rev. B {\bf 61}, 10267 (2000).
\bibitem{bocquet}
M. Bocquet, D. Serban, and M. R. Zirnbauer, Nucl. Phys. B {\bf 578}, 628 (2000).
\bibitem{chalker}
J. T. Chalker, N. Read, V. Kagalovsky, B. Horovitz, Y. Avishai, and A. W. W. Ludwig, cond-mat/0009463 (2000).
\bibitem{motrunich}
O. Motrunich, K. Damle, and D. A. Huse, cond-mat/0011200 (2000).
\bibitem{zubitz}
J. B. Zuber and C. Itzykson, Phys. Rev. D \textbf{15}, 2875 (1977).
\bibitem{benettin}
G. Benettin, L. Galgani, A. Giorgilli and J.-M. Strelcyn, Meccanica \textbf{15}, 9-30 (1980).
\bibitem{pichard}
J. L. Pichard and G. Sarma, J. Phys C {\bf 17}, 4111 (1981).
\bibitem{KrMcKi}
A. MacKinnon and B. Kramer, Phys. Rev. Lett. {\bf 47},
1546 (1981); Z. Phys. B {\bf 53}, 1 (1983).
\bibitem{kadce}
L. P. Kadanoff and H. Ceva, Phys. Rev. B \textbf{3}, 3918 (1971).
\bibitem{domany}
E. Domany, J. Phys. C \textbf{12}, L119 (1979).
\bibitem{eastmondhuckestein}
J. T. Chalker and J. Eastmond, unpublished; J. Eastmond, D.Phil. thesis, Oxford University (1992); B. Huckestein, Phys. Rev. Lett. {\bf 72}, 713 (1994).
\bibitem{binderyoung}
A. P. Young and K. Binder, Rev. Mod. Phys. {\bf 58}, 801 (1986).
\bibitem{bray}
A. J. Bray, private communication.
\bibitem{braymoore}
A. J. Bray and M. A. Moore, J. Phys. C{\bf 17} L463 (1984).
\bibitem{husemorg}
D. A. Huse and I. Morgenstern, Phys. Rev. B {\bf 32}, 3032 (1985).
\end{thebibliography}
\end{document}